\documentclass[reprint,longbibliography,
 amsmath,amssymb,superscriptaddress]{revtex4-2}
\usepackage{graphicx} 
\usepackage{dcolumn}
\usepackage{bm}
\usepackage[colorlinks=true, citecolor=blue, linkcolor=blue, urlcolor=blue]{hyperref}
\usepackage{xcolor}
\usepackage{amsmath, amssymb, amsfonts}
\usepackage{mathtools}
\usepackage{subfigure}
\usepackage{mathrsfs}
\usepackage{sidecap}
\usepackage{verbatim}

\definecolor{purple1}{rgb}{128,0,128}

\usepackage{bigints}
\newcommand{\bea}{\begin{eqnarray}}
\newcommand{\ea}{\end{eqnarray}}

\definecolor{darkpastelgreen}{rgb}{0.01, 0.75, 0.24}

\begin{document}

\title{Classical and quantum metrology of the Lieb-Liniger model}

\author{Jae-Gyun Baak}
\author{Uwe R. Fischer}
\affiliation{
 Seoul National University, Department of Physics and Astronomy,\\
 Center for Theoretical Physics, Seoul 08826, Korea
}

\date{\today}

\begin{abstract}
We study the classical and quantum Fisher information for the Lieb-Liniger model.
The Fisher information has been studied extensively
when the parameter is inscribed on a quantum state by a unitary process,
e.g., Mach-Zehnder or Ramsey interferometry.
Here, we investigate the case that
a Hamiltonian parameter to be estimated is imprinted on eigenstates of that Hamiltonian,
and thus is not necessarily encoded by a unitary operator.
Taking advantage of the fact that the Lieb-Liniger model is exactly solvable,
the Fisher information is determined for periodic and hard-wall boundary conditions,
varying number of particles, and for excited states of type-I and type-II in the Lieb-Liniger terminology.
We discuss the dependence of the Fisher information on interaction strength
and system size, to further evaluate the metrological aspects of the model.
Particularly noteworthy is the fact that
the Fisher information displays a maximum when we vary the system size,
indicating that the distinguishability of the wavefunctions is largest
when the Lieb-Liniger parameter is at the crossover
between Bose-Einstein condensate and Tonks-Girardeau limits.
The saturability of this Fisher information by the absorption imaging method is assessed
by a specific modeling of the latter.
\end{abstract}


\maketitle

\section{\label{sec:intro}Introduction}

The Lieb-Liniger (LL) model is the archetype of an exactly solvable many-body problem
with direct experimental relevance \cite{Paredes_2004,Kinoshita_2006}.  
It describes $N$ bosons interacting
via a repulsive contact interaction in one-dimensional space.
The time-independent Schrödinger equation with the LL Hamiltonian
has been solved by the method of coordinate Bethe ansatz \cite{Bethe_1931,Batchelor_2006}
with varying boundary conditions \cite{Lieb_Liniger_1963,Gaudin_1971}:
periodic and hard-wall boundary conditions,
which correspond to the particles moving along a ring and in a flat box, respectively.
The characteristic behavior of the energy spectrum
and the eigenstates of this model has been studied over the years in many guises;
see, for example, Refs.~\cite{Lieb_1963,Gaudin_1971,Batchelor_2005,Oelkers_2006,Gaudin_2014,Reichert_2019}.
The theoretical analyses of many-body systems have focused on special regimes of parameters
such as thermodynamic limit, weakly and strongly interacting regimes
due to their relative ease of capturing the physical features inherent to the systems
\cite{Dunjko_2001,Lang_2017,Minguzzi_ArXiv_2022}.
One notable case is that
the type-II elementary excitations in the LL model have been revealed to have correspondence to
the dark solitons of the Gross-Pitaevskii equation in the weak interaction limit
\cite{Kulish_1976,Ishikawa_1980,Mottelson_1999,Syrwid_2015,Golletz_2020,Syrwid_2021}.
On the other hand, there have been trials,
often using the multiconfigurational Hartree method,
to approach the regime
where the interaction strength is intermediate between its two extremes
or only a few particles are involved without thermodynamic limit
\cite{Dunjko_2001,Alon_2005,Marchukov_2019,Gwak_2021}.
Here, we focus on the Fisher information for the interaction strength
calculated with respect to the LL eigenstates with a few particles.

The Fisher information indicates the distinguishability of a physical state
by some measurement when a state parameter slightly changes.
It determines a minimum distance in the parameter space for different states to be resolved,
and only if the change of parameter value is larger than this distance,
one can detect the variation of state by the corresponding measurement
\cite{Wootters_1981}.
Thus the classical Fisher information (CFI) for a given measurement
defines the precision limit of estimating a parameter imprinted on a state,
and the quantum Fisher information (QFI) is the maximized CFI
when an optimal measurement is assumed.
All of the above is succinctly formulated by the Cramér-Rao inequality
\cite{Helstrom_1967,Braunstein_1994,Paris_2009,Wiseman_Milburn_2009,Giovannetti_2011}:
\begin{equation}
\big\langle(\delta\theta_\text{est})^2\big\rangle_\theta\ge\frac{1}{MF(\theta)}
\ge\frac{1}{M\mathfrak{F}(\theta)}\,,
\label{ineq:cramér-rao}
\end{equation}
where $\theta_\text{est}$ is an estimator
mapping $M$ measurement outcomes to an estimate for a parameter $\theta$,
and $\delta\theta_\text{est}$ is the bias-corrected deviation from the true value of $\theta$:
$\theta_\text{est}/|d\langle\theta_\text{est}\rangle_\theta/d\theta|-\theta$.
The second inequality, which provides the tighter bound,
represents the quantum version of the first one which gives the classical Cramér-Rao bound.
The expectation value of $(\delta\theta_\text{est})^2$ with respect to a state
with $\theta$ as a true value is denoted by $\langle(\delta\theta_\text{est})^2\rangle_\theta$.
Also, $F(\theta)$ and $\mathfrak{F}(\theta)$ represent
the CFI and the QFI for the estimation of $\theta$, respectively.
The first inequality in Eq.~\eqref{ineq:cramér-rao} is always
asymptotically, i.e., as $M\rightarrow\infty$, saturable
by using the maximum likelihood estimator (MLE) for $\theta_\text{est}$,
and the second inequality can always be saturated theoretically
by finding the optimal measurement \cite{Braunstein_1994,Paris_2009}.

Quantum parameter estimation theory concerns
estimating a parameter $\theta$ in a Hamiltonian $H_\theta$
and suggests the general measurement scheme consisting of four steps \cite{Giovannetti_2006}:
1) the preparation of an initial quantum state, say $\rho_0$,
2) the imprint of a parameter on the state during its physical process,
where typically $\rho_\theta=U_\theta\rho_0U_\theta^\dagger$
with a unitary operator $U_\theta$ derived from $H_\theta$,
3) the measurement of the final state $\rho_\theta$, and after repeating these three steps
$M$ times, 4) the estimation of the parameter $\theta$ from those $M$ measurement results.
Many studies have tried to improve the $N$-scaling of QFI in this protocol,
where $N$ is the number of quantum resource like equally prepared quantum states,
basically trying to surpass the standard quantum (shot noise) limit, $\mathfrak{F}(\theta)\sim N$,
and ultimately achieve the Heisenberg limit, $\mathfrak{F}(\theta)\sim N^2$, whether this is possible.
The enhancement of $N$-scaling means the better measurement precision
with the same amount of resource ($N$ copies of some fundamental entity), and can be theoretically accomplished
by preparing special initial states, which are commonly entangled or squeezed,
in step 1 \cite{Braunstein_1992,Giovannetti_2004,Boixo_2007,Giovannetti_2011}
or utilizing various dynamics during the state evolution
in step 2 \cite{Boixo_2008_01,Boixo_2008_07,Pang_2017}.
Such studies that concentrate on the QFI assume the saturability of $F(\theta)\le\mathfrak{F}(\theta)$
by choosing the optimal measurements in step 3.
However, these theoretically obtained measurements are usually distant
from those implemented in the real experiments
except for a few cases like spin measurement in the magnetic field \cite{Barndorff-Nielsen_2000}.
Thus the actual precision is often evaluated by the CFI
on the basis of the realistic measurements
such as the measurement of population imbalance between two possible outputs
\cite{Gietka_2014,Wasak_2016}.

Quantum metrological framework so far has mainly been discussed
within the SU(2) representation, which mathematically describes the two-state systems
like two polarization states of a photon, two spin states of a spin-$1/2$ atom,
and two locations of an atom in a double-well potential.
Then two-mode interferometric methodology, e.g. Mach-Zehnder or Ramsey interferometry,
makes use of the interference
between two internal states or two spatially separated states
in order to extract the relative phase between the two states,
from which one can estimate the target parameters.
Particularly, the quantum metrology
with the ultracold bosonic gas such as Bose-Einstein condensates (BECs)
in a double-well trap has been vastly studied
\cite{Dalton_2007,Grond_2011,Dalton_2012,Juha_2012,Juha_2014,Gietka_2014}.
Such metrology using the trapped atomic systems depends on the validity of two-mode approximation,
the conditions of which include small interparticle interaction
or sufficient separation of the two wells in a double-well potential \cite{Milburn_1997}.
Even in the case of bosons in a single harmonic trap,
the SU(2) metrology can be applied
if one can assume the regimes of parameters
where the two-mode approximation effectively describes the dynamics of the systems
\cite{Julien_2019}.

In this paper, we analyze the metrological scenario
of estimating a Hamiltonian parameter
using the eigenstates of that Hamiltonian,
which naturally contain the parameter to be estimated in a non-unitary way.
This scenario is involved in a single state,
especially in the position representation,
thus the interference between any two states is not taken into account,
whereas in the SU(2) scenario
the superposition of two orthogonal states is used to produce
the relative phase originating from the distinct evolutions of those states under a unitary process.
The estimation efficiency is evaluated
by the Fisher information defined in terms of the wavefunction,
and we concentrate on the analysis of the wavefunction-based Fisher information itself
rather than its $N$-scaling.
All of these are discussed with the LL model,
where the position measurement of the particles is chosen
to calculate the CFI and evaluate the precision of estimating the interaction strength.
Instead of finding the optimal measurements,
we discuss the conditions under which the position measurement is optimal
and try to assess the realistic measurement precision
based on the absorption imaging of atomic cloud,
which is shown to be the imperfect version of the position measurement.

This paper is organized as follows.
In Sec.~\ref{sec:model}
we look through the LL model including the eigenfunctions and their norms,
which are necessary to calculate the Fisher information,
under two different boundary conditions: periodic and hard-wall boundary conditions.
In Sec.~\ref{sec:fisherinfo}
we first introduce the appropriate forms of CFI and QFI in terms of the wavefunction
and relate the saturability of $F(\theta)\le\mathfrak{F}(\theta)$ to certain class of wavefunctions.
Then we construct a mathematical model of the imperfect measurement of the particle positions
in order to simulate the absorption imaging method,
which is the main measurement tool to explore the ultracold atomic systems
\cite{Anderson_1995,Andrews_1997}.
Lastly we develop the formula of Fisher information for the interaction strength
calculated with respect to the LL eigenfunctions
and discuss its functional properties.
Also, the issue of computing the scalar product between two LL eigenstates
that one encounters while calculating the Fisher information is explained.
In Sec.~\ref{sec:analysis}
the Fisher information thus obtained is plotted
versus the interaction strength or the system size,
and several notable features are discussed.
We consider the effect of the number of particles
and the types of elementary excitations on the Fisher information.
In every case, the Fisher information values for two different boundary conditions are compared.
Finally we show that the precision limit defined by the Fisher information
can be achieved by the absorption imaging as its resolution improves.

\section{\label{sec:model}Lieb-Liniger Model Under Different Boundary Conditions}

The one-dimensional dynamics of $N$ bosons
interacting via a repulsive $\delta$-function potential
is described by the LL Hamiltonian
\begin{equation}
\hat{H}_\text{LL}=-\sum_{j=1}^N\frac{\partial^2}{\partial x_j^2}+2\,c\sum_{j<l}\delta(x_j-x_l)\,,
\end{equation}
where $c>0$ due to the repulsion and $\hbar=2m=1$ for the units.
The time-independent Schrödinger equation, $\hat{H}_\text{LL}\psi=E\,\psi$, is solved
by the Bethe method with appropriate boundary conditions \cite{Lieb_Liniger_1963,Gaudin_1971}.
The process of Bethe method is the following:
1) Under the assumption that any two particles do not cross each other,
each particle is regarded as a free particle.
Hence a functional form called \textit{Bethe ansatz} is given by the superposition of
free particle wavefunctions in the domain $D:x_{1}<x_{2}<\cdots<x_{N}$.
2) The $\delta$-function potential gives a continuity equation,
i.e., $[\partial\,\psi/\partial\,x_{j+1}-\partial\,\psi/\partial\,x_j=2\,c\,\psi]|_{x_{j+1}=x_j}$,
determining the form of coefficients in the Bethe ansatz.
3) Applying the boundary condition leads to the \textit{Bethe equations}
which determine the \textit{quasi-momenta} introduced by the free particle wavefunctions.
The $N$-boson wavefunctions must be symmetric under the exchange of arbitrary two spatial coordinates,
thus the Bethe ansatz defined only in the domain $D$ so far covers the entire space $[0,L]^N$.
Now we have the exact LL eigenfunctions.
4) In addition to these, however, we devote our particular attention to the norm of Bethe ansatz,
hence we explicitly represent the norm of LL eigenfunction under a given boundary condition.

\begin{figure}[t]
\includegraphics[width=\columnwidth]{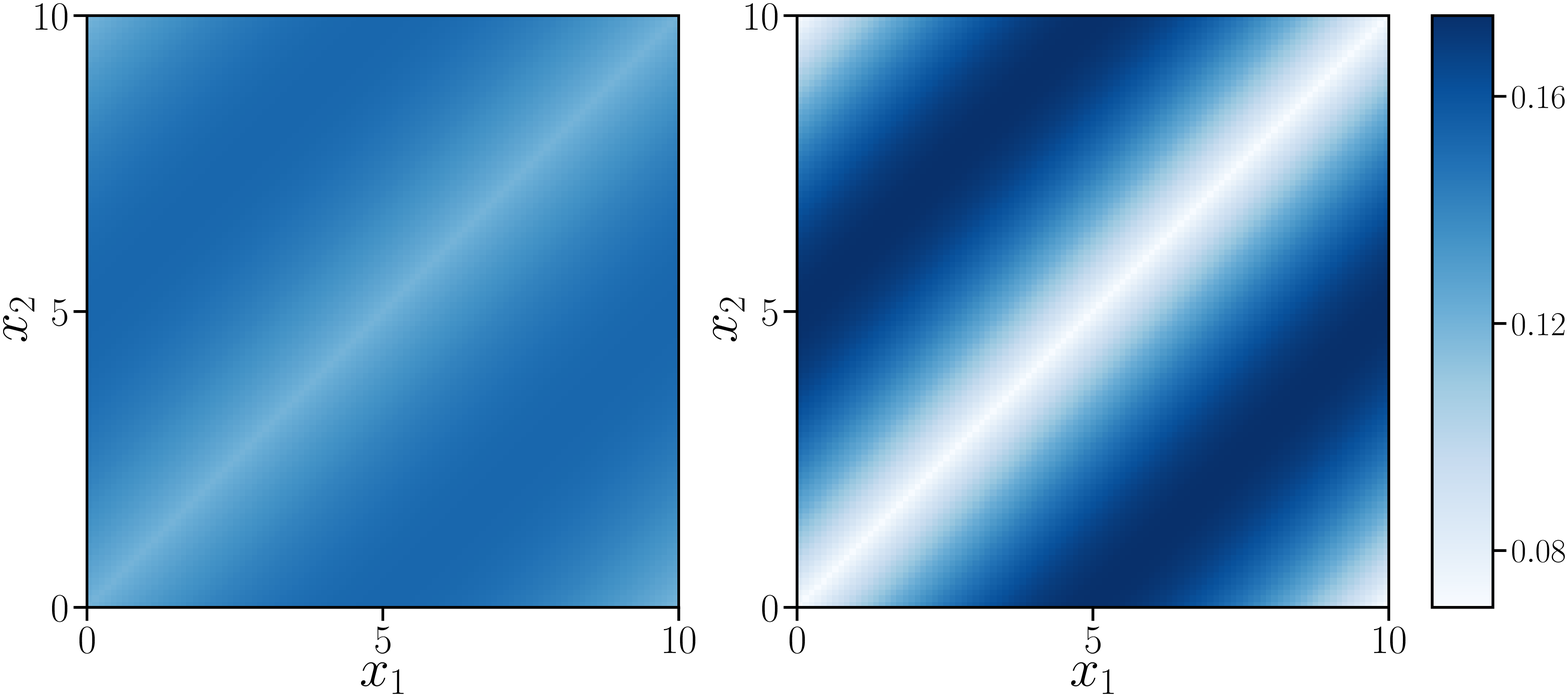}
\includegraphics[width=\columnwidth]{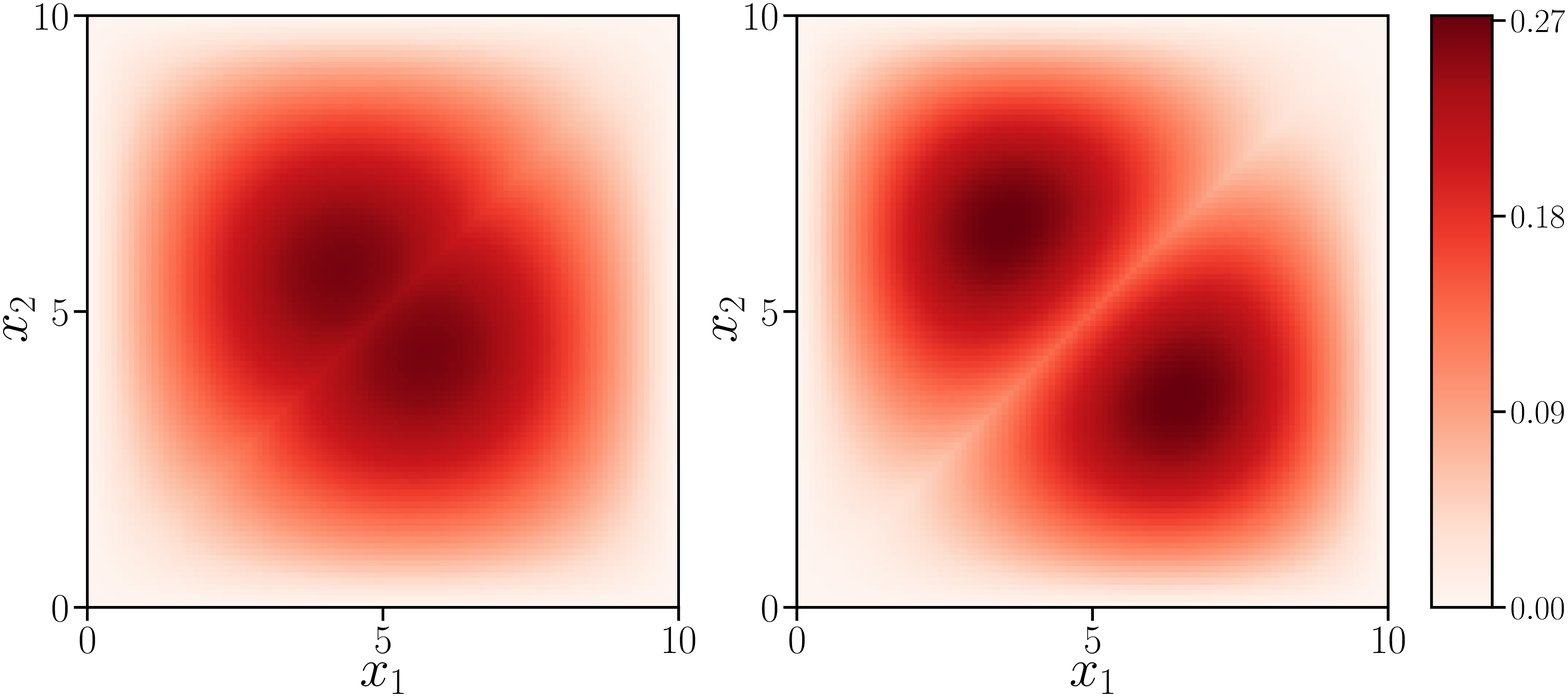}
\caption{\label{fig:grwf}
The normalized ground-state wavefunctions of $N=2$ particles
in the LL model under periodic boundary condition (upper row)
and hard-wall boundary condition (lower row)
with the increasing interaction strength $c$ from left to right in each row.
One can see how different geometries defined by boundary conditions
are reflected in the wavefunction profiles.
When $c$ is small, the two particles are more likely to share a common space,
while they move away from each other as $c$ increases.
Between the two extreme regimes of $c$, i.e., $c=0$ and $c=\infty$,
there appears a maximum sensitivity of the wavefunction with respect to $c$,
which is reflected in the larger Fisher information in that region which we reveal below.}
\end{figure}

The most well-known solution of the LL model is the one under the periodic boundary condition
describing the one-dimensional system of $N$ bosons moving along a ring.
The Bethe ansatz for this boundary condition is given by
\begin{equation}
\tilde{\psi}(x_1,\cdots,x_N)=\sum_{P's}A(P)\,e^{\,i\sum_{j=1}^Nk_{P_j}x_j}\,,
\label{eq:periodicLLsol}
\end{equation}
where the continuity equation gives
\[
A(P)=\prod_{j<l}\bigg(1+\frac{ic}{k_{P_j}-k_{P_l}}\bigg).
\]
The tilde of $\tilde{\psi}$ in Eq.~\eqref{eq:periodicLLsol} emphasizes that it is unnormalized.
The sum over $P$'s means the sum over a set of the permutations of the integers from 1 to $N$.
The energy and the momentum of the state described by this wavefunction are
$E_N=\sum_{j=1}^Nk_j^2$ and $P_N=\sum_{j=1}^Nk_j$, respectively.
Then applying the periodic boundary condition introduces the system size $L$
by imposing the cyclicity condition on the wavefunction,
i.e., $\tilde{\psi}(x_1,\cdots,x_N)=\tilde{\psi}(x_2,\cdots,x_N,x_1+L)$,
which is now defined in the domain $D':0<x_1<\cdots<x_N<L$.
From this condition, one obtains a system of $N$ equations, i.e., Bethe equations:
\begin{equation}
Lk_j=2\pi I_j-2\sum_{l=1}^N\tan^{-1}\!\Big(\frac{k_j-k_l}{c}\Big),
\label{eq:periodicBetheeq}
\end{equation}
which relates the quasi-momenta $k_j$'s to the values of $I_{j}$'s given $c$ and $L$.
The quantum numbers, $I_1<I_2<\cdots<I_N$, are half-integers for even $N$ or integers for odd $N$
and do not allow duplicate values.
The symmetric distribution of $I_j$'s around $0$ with interval $1$,
i.e., $[I_1=-(N-1)/2,I_2=-(N-3)/2,\cdots,I_N=(N-1)/2]$,
gives the symmetric distribution of $k_j$'s around $0$,
which obviously minimizes the energy.
Thus it corresponds to the ground state with minimal energy and zero momentum.
The norm of the unnormalized wavefuction in Eq.~\eqref{eq:periodicLLsol}
has been conjectured in Ref.~\cite{Gaudin_2014} and proved in Ref.~\cite{Korepin_1982}
using the method of algebraic Bethe ansatz.
The norm squared of Eq.~\eqref{eq:periodicLLsol} has thus been established to be
\begin{equation}
\mathcal{N}^{\,2}=\prod_{j<l}\Big(1+\frac{c^2}{(k_j-k_l)^2}\Big)\,\text{det}[\mathbf{H}]\,,
\label{eq:periodicLLnorm}
\end{equation}
where the matrix $\mathbf{H}$ is
\[
[\mathbf{H}]_{ij}=\delta_{ij}\Big(L+\sum_{l=1}^N\frac{2c}{(k_i-k_l)^2+c^2}\Big)-\frac{2c}{(k_i-k_j)^2+c^2}\,.
\]

The LL model with the hard-wall boundary condition
has been covered in Refs.~\cite{Gaudin_1971,Batchelor_2005,Oelkers_2006,Gaudin_2014,Reichert_2019}.
The Bethe ansatz for this boundary condition is given by
\begin{equation}
\tilde{\psi}(x_1,\cdots,x_N)=\sum_{\epsilon's}\sum_{P's}\,\pi_\epsilon\,
A(\epsilon,P)\,e^{i\sum_{j=1}^N\epsilon_jk_{P_j}x_j}\,,
\label{eq:hardwallLLsol}
\end{equation}
where the continuity equation gives
\begin{equation}
A(\epsilon,P)=\prod_{j<l}\Big[1-\frac{i\,c}{\epsilon_jk_{P_j}+\epsilon_lk_{P_l}}\Big]
\Big[1+\frac{i\,c}{\epsilon_jk_{P_j}-\epsilon_lk_{P_l}}\Big].\nonumber
\end{equation}
Here, $\epsilon:=\{\epsilon_1,\cdots,\epsilon_N\}$
and each $\epsilon_j$ is summed over $\pm1$.
The $P$ runs over all permutations of $[1,2,\cdots,N]$ as in Eq.~\eqref{eq:periodicLLsol}
and $\pi_\epsilon$ is the product of all $\epsilon_j$'s: $\epsilon_1\epsilon_2\cdots\epsilon_N$.
The sign of $k_j$ is absorbed into $\epsilon_j$ to make $k_j>0$ for all $j$'s
and $k_1<k_2<\cdots<k_N$ is assumed without loss of generality due to the sum over $P$.
The application of the hard-wall boundary condition introduces the system size $L$
by imposing a condition on the wavefunction,
i.e., $\tilde{\psi}(x_1=0,x_2,\cdots,x_N)=\tilde{\psi}(x1,\cdots,x_{N-1},x_N=L)=0$,
which is now defined in the domain $D'$.
This condition finally gives the Bethe equations for quasi-momenta:
\begin{equation}
Lk_j=\pi I_j-\sum_{l\neq j}\Big[\tan^{-1}\!\Big(\frac{k_j-k_l}{c}\Big)
+\tan^{-1}\!\Big(\frac{k_j+k_l}{c}\Big)\Big].
\label{eq:hardwallBetheeq}
\end{equation}
The quantum numbers, $I_1<I_2<\cdots<I_N$, are positive integers
irrespective of the evenness of $N$ and do not allow duplicate values,
thus the ground state corresponds to $[I_1=1,I_2=2,\cdots,I_N=N]$,
which minimizes the energy $E_N=\sum_{j=1}^Nk_j^2$.
Applying the conjecture about the norm of Bethe ansatz in Ref.~\cite{Gaudin_2014}
to Eq.~\eqref{eq:hardwallBetheeq} leads to
the norm squared of Eq.~\eqref{eq:hardwallLLsol}
that is expressible as Eq.~\eqref{eq:hardwallnorm} in the Appendix.

The LL model under either of the boundary conditions above
is fully described by three parameters:
interaction strength $c$, system size $L$, and particle number $N$.
If one considers, however, the dimensionless form of LL model,
i.e., $k_j\rightarrow\tilde{k_j}=L\,k_j$, $c\rightarrow\tilde{c}=c\,L$,
and $x_j\rightarrow\tilde{x_j}=x_j/L$,
one can find that the system depends only on the two parameters: $c\,L$ and $N$.
Basically this origins from the characteristics of Bethe equations,
Eq.~\eqref{eq:periodicBetheeq} and Eq.~\eqref{eq:hardwallBetheeq},
that for fixed $N$ the Bethe equations are invariant under
any change of $c$ and $L$ keeping $c\,L$ constant.
Hence for a fixed number of particles,
the value of $\tilde{c}=c\,L$ completely determines the value of $\tilde{k}_j=Lk_j$
and thus the system.
In thermodynamic limit, i.e., $N\rightarrow\infty$ with $N/L$ fixed,
the LL model is completely described by a single parameter
called Lieb's parameter: $\gamma=c/\rho$,
where $\rho$ is the number density $N/L$.
Also, in this limit, the different effects between the two boundary conditions disappear
and even the properties of bosonic gas along the ring space can be considered
as if the gas resides in a flat box potential.

Though dimensionless form helps understand the essence of a system
irrespective of the unit choice for the parameters,
in order to test and apply the Fisher information
as a realistic indicator for the limit of measurement precision,
we don't take such dimensionless parameters
so that $c$ and $L$ can be treated separately.
By doing so, we can verify
how the independent adjustment of one parameter
can help improve the measurement precision of the other,
thus we can simulate the situations encountered in the real measurements.
Also, we don't assume thermodynamic limit and keep small number of particles
to investigate the role of different geometries
specified by different boundary conditions: periodic and hard-wall boundary conditions.
Hence the system is now described by three independent parameters:
$c$, $L$, and $N$.

\section{\label{sec:fisherinfo}Fisher Information in the Lieb-Liniger Model}

\subsection{\label{subsec:fisherinfo}Fisher information of the wavefunction}

The Fisher information has played an important role in the parameter estimation theory
in that it determines the precision to which one can fundamentally estimate a parameter
encoded in a probability distribution.
For any conditional probability density $P(X|\theta)$ of a continuous random variable $X$,
the CFI of a parameter $\theta$ is
\begin{equation}
F(\theta)=\int\!dx\,\frac{1}{P(x|\theta)}\Big(\frac{\partial P(x|\theta)}{\partial\theta}\Big)^{\!2}\,,
\label{eq:cfi1}
\end{equation}
where the integral (or sum for a discrete random variable) is done
over the domain of single measurement result $x$.
The $F(\theta)$ quantifies
how well two slightly different values of $\theta$ can be distinguished
based on a single measurement outcome $x$.
When a set of $M$ outcomes
from $M$ independently and identically repeated measurements is used
to estimate the $\theta$,
the inverse of $\sqrt{MF(\theta)}$ provides the lower limit of estimation error
as in Eq.~\eqref{ineq:cramér-rao}.

In quantum theory,
a quantum state parametrized by a parameter $\theta$ is connected
to a probability distribution of the measurement results by
$P(x|\theta)=\text{Tr}[\rho_\theta\hat{E}_x]$,
where $\rho_\theta$ is the quantum state
and a set of $\hat{E}_x$'s for all possible $x$'s, i.e., $\{\hat{E}_x:x\}$,
is the positive operator-valued measure (POVM)
specifying the type of measurement.
Given $\rho_\theta$,
the $\text{Tr}[\rho_\theta\hat{E}_x]$ allows additional optimization
over the POVMs in a way that $F(\theta)$ is maximized.
The maximal $F(\theta)$ defines the QFI of $\theta$:
\begin{equation}
\mathfrak{F}(\theta)=\max_{\{\hat{E}_x:x\}}\!F(\theta)=\text{Tr}[\rho_\theta L_\theta^2]\,,
\label{eq:qfi1}
\end{equation}
where $L_\theta$ is the symmetric logarithmic derivative satisfying
$\partial_\theta\rho_\theta=(\rho_\theta L_\theta+L_\theta\rho_\theta)/2$.
By Eq.~\eqref{ineq:cramér-rao}, the QFI specifies the precision
one can ultimately attain for estimating an unknown parameter
encoded in a quantum state.
Note that the optimal POVM making $F(\theta)$ reach its maximum can always be found.
When the parameter $\theta$ is encoded
by a unitary process with some generator $\hat{G}$,
i.e., $\rho_\theta=e^{-i\theta\hat{G}}\rho_0\,e^{i\theta\hat{G}}$,
the upper bound of QFI is given as $4\langle\psi_0|(\Delta G)^2|\psi_0\rangle$,
which obviously can be reached when $\rho_0$ is a pure state: $\rho_0=|\psi_0\rangle\langle\psi_0|$.
Thus, for general states including mixed ones,
$\mathfrak{F}(\theta)\leq4\langle\psi_0|(\Delta G)^2|\psi_0\rangle$ \cite{Wiseman_Milburn_2009}.

Here, however, we are concerned with the Fisher information of a Hamiltonian parameter
that a quantum state naturally has after being realized as an eigenstate of that Hamiltonian
rather than unitarily encoded.
For the generic pure state $|\psi_\theta\rangle$ with $\theta$
encoded via a physical process, the QFI defined in (\ref{eq:qfi1}) reduces to
\begin{eqnarray}
\!\!\!\!\!\!\!\!\!\!\mathfrak{F}(\theta)&=&4\,\Big(\langle\partial_\theta\psi_\theta|\partial_\theta\psi_\theta\rangle
-\big|\langle\psi_\theta|\partial_\theta\psi_\theta\rangle\big|^2\Big)\nonumber\\
&=&4\,\Big(\!\int\!\!d\mathbf{x}\,\big|\partial_\theta\psi_\theta(\mathbf{x})\big|^2
\!\!-\big|\!\!\int\!\!d\mathbf{x}\,\psi_\theta^\ast(\mathbf{x})\,\partial_\theta\psi_\theta(\mathbf{x})\big|^2\Big),
\label{eq:qfi2}
\end{eqnarray}
where it is also expressed in the position representation
with an $N$-particle position state $|\mathbf{x}\rangle:=|x_1,x_2,\cdots,x_N\rangle$
satisfying $\int\!d\mathbf{x}|\mathbf{x}\rangle\langle\mathbf{x}|=\hat{1}$
and $\langle\mathbf{x}|\mathbf{x}'\rangle=\delta(\mathbf{x}-\mathbf{x}')$.
When $\psi_\theta(\mathbf{x})$ is the most general complex-valued function,
we can consider the polar representation of a wavefunction:
$\psi_{\theta}(\mathbf{x})=|\psi_{\theta}(\mathbf{x})|e^{i\varphi_\theta(\mathbf{x})}$,
with some real-valued function $\varphi_\theta(\mathbf{x})$.
Then $\mathfrak{F}(\theta)$ in (\ref{eq:qfi2}) can be further tranformed into
\begin{eqnarray}
\frac{\mathfrak{F}(\theta)}4
=&&\int\!\!d\mathbf{x}\,\big(\partial_\theta |\psi_\theta(\mathbf{x})|\big)^2
+\int\!\!d\mathbf{x}\,|\psi_\theta(\mathbf{x})|^2\big(\partial_\theta\varphi_\theta(\mathbf{x})\big)^2\nonumber\\
&&\qquad\qquad\qquad
-\,\,\Big(\!\int\!\!d\mathbf{x}\,|\psi_\theta(\mathbf{x})|^2\,\partial_\theta\varphi_\theta(\mathbf{x})\Big)^{\!2},
\label{eq:qfi3}
\end{eqnarray}
where $\int\!d\mathbf{x}\,|\psi_\theta(\mathbf{x})|\,\partial_\theta|\psi_\theta(\mathbf{x})|=0$ derived
from the $\theta$-derivative of $\int\!d\mathbf{x}\,|\psi_\theta(\mathbf{x})|^{2}=1$ is used.
The second and third terms in (\ref{eq:qfi3}) combine to be the variance of $\partial_\theta\varphi_\theta(\mathbf{x})$
with respect to the probability distribution $P(\mathbf{x}|\theta)=|\psi_\theta(\mathbf{x})|^2$.
On the other hand, the CFI in (\ref{eq:cfi1}) can be recast in terms of the wavefunction
by the relation $P(\mathbf{x}|\theta)=|\psi_\theta(\mathbf{x})|^2$:
\begin{equation}
F(\theta)=4\!\int\!d\mathbf{x}\,\big(\partial_\theta|\psi_\theta(\mathbf{x})|\big)^{\!2}\,,
\label{eq:cfi2}
\end{equation}
and finally it obviously follows
that the QFI in Eq.~\eqref{eq:qfi3} comprises two parts \cite{Shun-Long_2006,Facchi_2010}:
\begin{equation}
\mathfrak{F}(\theta)=F(\theta)+4\,\text{Var}\big[\partial_\theta\varphi_\theta\big]_{\psi_\theta}\,,
\label{eq:qfi4}
\end{equation}
in which $\text{Var}[A]_{\psi_\theta}$ means the variance of a quantity $A(\mathbf{x})$
with respect to the probability distribution $|\psi_\theta(\mathbf{x})|^2$.

To represent the CFI in terms of the wavefunction implies
that a set of $\hat{E}_\mathbf{x}=|\mathbf{x}\rangle\langle\mathbf{x}|$,
i.e., $\{|\mathbf{x}\rangle\langle\mathbf{x}|:\mathbf{x}\}$, is chosen as a POVM,
which means the projective measurement of the positions of $N$ particles.
This POVM is not optimal according to Eq.~\eqref{eq:qfi4},
but we can find it optimal under certain conditions.
One notable case is when $\psi_\theta(\mathbf{x})=e^{i\varphi(\theta)}\phi_\theta(\mathbf{x})$
with $\varphi(\theta)$ and $\phi_\theta(\mathbf{x})$ being real-valued.
Since $\varphi(\theta)$ is independent of $\mathbf{x}$,
the variance term in Eq.~\eqref{eq:qfi4} disappears,
hence $\mathfrak{F}(\theta)=4\int d\mathbf{x}\,(\partial_\theta\phi_\theta(\mathbf{x}))^2=F(\theta)$.
For example, if $\psi_\theta(\mathbf{x})$ is real-valued, i.e., $\varphi(\theta)=0$,
or is purely imaginary-valued, i.e., $\varphi(\theta)=\pi/2$,
the QFI will be $4\int d\mathbf{x}\,(\partial_\theta\psi_\theta(\mathbf{x}))^2$
or $-4\int d\mathbf{x}\,(\partial_\theta\psi_\theta(\mathbf{x}))^2$, respectively,
which is exactly equal to the CFIs
calculated by Eq.~\eqref{eq:cfi2} with the corresponding $\psi_\theta(\mathbf{x})$.
If $\phi_\theta(\mathbf{x})$ is an eigenstate of a Hamiltonian,
which can always be chosen to be real,
then $e^{i\varphi(\theta)}\phi_\theta(\mathbf{x})$ is a representation
of its time evolution, where $\varphi_\theta$ contains the eigenenergy
of the state that also relies on the Hamiltonian parameter $\theta$.
Thus the saturation of CFI to QFI is maintained along the time evolution.
Finally, the Fisher information defined as above applies to the case when
one tries to estimate a Hamiltonian parameter $\theta$
from the position measurement on the eigenstates of that Hamiltonian.

\subsection{\label{subsec:measure}Single-shot measurement with finite resolution}

The ultimate quantum limit of any measurement precision is identified by the QFI,
but the feasible limit is always given by the CFI, that is, by a specific choice of measurement.
Relating the probability density $P(x|\theta)$ to the wavefunction $\psi_\theta(x)$ introduces a (hypothetical) measurement
where the positions of $N$ particles are determined with infinite level of precision.
Even though we have discussed the condition on $\psi_\theta(x)$ to make
$\mathfrak{F}(\theta)=F(\theta)$, what one actually encounters in the real experiments
is the imperfect measurement allowing for uncertainty in the measured positions,
which can be supported by the absorption imaging method
that has been used as a measurement to investigate the cold atomic systems
\cite{Dalfovo_1999}.
This method is implemented by irradiating laser light from above to the cold atomic gas
and measuring the brightness captured by the CCD camera installed below the gas.
The brightness profile generated from the camera is fitted
to obtain the column density distribution of the gas,
hence this absorption imaging is thought of as a measurement of the number of particles
in each pixel area of the camera.
Then one can estimate any system parameter from the density profile.
By using a model of single-shot measurement with finite resolution to simulate the absorption imaging
\cite{Sakmann_2016,Lode_2017,Lode_2021},
it can be shown that as narrowing the pixel size, one can asymptotically obtain
the precision level provided by the perfect position measurement.

The imperfect measurement of the position of a single particle in one-dimensional space
with finite resolution $\Delta x_p$ is described by the POVM $\{\hat{\Pi}_j:j\}$ satisfying
\begin{equation}
\hat{\Pi}_j:=\!\int_{a_j}^{a_{j+1}}\!\!\!\!\!\!\!\!dx\,\hat{\Pi}(x)\,,\,\,\,\,
\sum_j\hat{\Pi}_j=\!\int_{-\infty}^\infty\!\!\!\!dx\,\hat{\Pi}(x)=\mathbb{I}\,,
\label{eq:imperfectpovm}
\end{equation}
where $\hat{\Pi}(x)=|x\rangle\langle x|$ forms a POVM for the perfect position measurement
and $a_{j+1}-a_j=\Delta x_p$ for all $j$'s.
It is easy to show that $\hat{\Pi}_j\,\hat{\Pi}_l=\delta_{jl}$,
which is the property satisfied by the projective measurements.
Thus the probability of finding a particle within a range $(a_j,a_{j+1})$,
i.e., $\text{Tr}[\rho_\theta\hat{\Pi}_j]$, is considered
instead of the one of finding the particle exactly at $x$,
i.e., $\text{Tr}[\rho_\theta\hat{\Pi}(x)]dx$,
for a given quantum state $\rho_\theta$.

We extend this consideration to the case of $N$ particles and $N_p$ pixels.
The coordinate range of $j$th pixel is denoted by $A_j:=(a_{j-1},a_j)$,
where $j=0,1,\cdots,N_p,N_p+1$ and $a_j-a_{j-1}=\Delta x_p$ for all $j$'s except for $0$ and $N_p+1$.
Also, we define $a_{-1}:=-\infty$ and $a_{N_p+1}:=\infty$.
A one-dimensional absorption image can be written as
$\mathfrak{n}:=(\mathfrak{n}_0,\mathfrak{n}_1,\cdots,\mathfrak{n}_{N_p},\mathfrak{n}_{N_p+1})$,
where $\mathfrak{n}_{j\neq0,N_p+1}$ is the number of atoms found in $j$th pixel bin
and $\sum_{j=0}^{N_p+1}\mathfrak{n}_j=N$, i.e., the total number of particles.
We assume the length $\Delta x_{p}N_{p}$ should be large enough to cover the whole system,
but the numbers of atoms outside the measured area,
$\mathfrak{n}_0$ for the left and $\mathfrak{n}_{N_p+1}$ for the right,
are considered for theoretical completion.

Now one can think of the probability distribution of absorption images:
\begin{equation}
P(\mathfrak{n}|\theta)=\zeta_\mathfrak{n}\!\!\int_A\!d\mathbf{x}\,P(\mathbf{x}|\theta)\,,
\sum_{\mathfrak{n}\in H(N,N_p)}\!\!\!\!\!\!P(\mathfrak{n}|\theta)=1\,,
\label{eq:singleshotprob}
\end{equation}
where $P(\mathbf{x}|\theta)$ is obtained from the absolute-square of $N$-particle wavefunction,
i.e., $|\psi_\theta(\mathbf{x})|^2$, and
\[
\zeta_\mathfrak{n}:=\frac{N!}{\prod_{j=0}^{N_p+1}\mathfrak{n}_j!}\,,\,\,
\int_{A}d\mathbf{x}:=\int_{A_{j_1}}\!\!\!\!dx_1\int_{A_{j_2}}\!\!\!\!dx_2\,\cdots\int_{A_{j_N}}\!\!\!\!\!\!dx_N\,.
\]
Also, $H(N,N_{p})$ is a set of the different ways
of selecting $N$ elements out of $\{A_j:j=0,1,\cdots,N_p+1\}$ allowing for duplication,
i.e., a set of all absorption images that can be realized.
The integration part gives the probability of finding the 1st particle in the range $A_{j_1}$,
the 2nd particle in the range $A_{j_2}$, and so on.
Note that the set $A=\{A_{j_1},\cdots,A_{j_N}\}$ is in an ascending order,
i.e., $j_1\le j_2\le\cdots\le j_N$, and that's why $\zeta_\mathfrak{n}$ is multiplied
to consider the indistinguishability of particles.
In short, the $A$ is an area in the $N$-dimensional space of $(x_{1},\cdots,x_{N})$,
where any element leads to the same absorption image, i.e., $\mathfrak{n}$.
Finally, using Eq.~\eqref{eq:singleshotprob},
the CFI for the absorption imaging is obtained as 
\begin{equation}
F(\theta)\quad=\sum_{\mathfrak{n}\in H(N,N_p)}\frac{1}{P(\mathfrak{n}|\theta)}\Big(\frac{dP(\mathfrak{n}|\theta)}{d\theta}\Big)^2\,,
\label{eq:imperfectcfi}
\end{equation}
which determines the precision level for estimating $\theta$ of $\rho_\theta$
from the results of absorption imaging method.

For example, let us suppose $N=10$ and $N_{p}=3$,
that is, 10 particles and 3 pixels, thus a case of very low resolution.
Then the integration area, i.e., $A$, corresponding to an absorption image $\mathfrak{n}=(1,5,1,3,0)$ will be
$\{A_{0},A_{1},A_{1},A_{1},A_{1},A_{1},A_{2},A_{3},A_{3},A_{3}\}$
and $\zeta_{\mathfrak{n}}=10!/(1!5!1!3!0!)=5040$.
Like this, any absorption image has a corresponding representation of $A$.
The probability of attaining $(1,5,1,3,0)$ as an absorption image can be calculated as
\[
5040\times\!\!\int_{-\infty}^{a_0}\!\!\!\!dx_1\int_{a_0}^{a_1}\!\!\!\!dx_2\int_{a_0}^{a_1}\!\!\!\!dx_3
\cdots\!\!\int_{a_2}^{a_3}\!\!\!\!dx_{10}\,P(\mathbf{x}|\theta)\,,
\]
where $P(\mathbf{x}|\theta)=|\psi_\theta(\mathbf{x})|^2$
and the integral about each range is implemented with the repetition indicated by $(1,5,1,3,0)$.
Also, $H(N,N_{p})$ is a set of all absorption images
that can be realized by $10$ particles with $3+2$ ranges:
\[
\{(10,0,0,0,0),(9,1,0,0,0),\cdots(0,0,0,0,10)\}
\]
with the total number of elements $\frac{(N+N_{p}+1)!}{(N_{p}+1)!N!}=1001$.
Now we have everything in hand to calculate the CFI for the absorption imaging.
Later on, we will check by numerical simulation that the CFI of the single-shot measurement with finite resolution,
i.e., Eq.~\eqref{eq:imperfectcfi},
converges to the CFI provided by the wavefuction, i.e., Eq.~\eqref{eq:cfi2},
as narrowing $\Delta x_p$.

\subsection{\label{subsec:}Fisher information of the Bethe ansatz}

We suppose that the interaction strength $c$ is a target parameter
to be estimated and the system size $L$ is a known resource parameter that
we can directly control in the LL model. The $L$ is a resource in a sense that
it serves a possibility of enhancing measurement precision of another parameter $c$.
Note that we focus on the estimation of a Hamiltonian parameter
that is encoded into a state while it is being realized as an eigenstate of the Hamiltonian
rather than by evolving through a unitary evolution,
e.g., Mach-Zehnder or Ramsey interferometry.
Hence we use Eq.~\eqref{eq:qfi2} and Eq.~\eqref{eq:cfi2} in Sec.~\ref{subsec:fisherinfo}
to calculate the QFI and the CFI directly from the wavefunctions.
Since the wavefunctions in Eq.~\eqref{eq:periodicLLsol} and Eq.~\eqref{eq:hardwallLLsol} are unnormalized,
the expressions of QFI and CFI, i.e., Eq.~\eqref{eq:qfi2} and Eq.~\eqref{eq:cfi2},
need to be rewritten in terms of unnormalized wavefunctions:
\begin{subequations}
\begin{eqnarray}
\mathfrak{F}=\frac{4}{\mathcal{N}^2}\bigg[\!
\int\!d\mathbf{x}\,\Big|\frac{d\tilde{\psi}}{dc}\Big|^2\!\!
-\frac{1}{\mathcal{N}^2}\Big|\!\int\!d\mathbf{x}\,\tilde{\psi}^\ast\frac{d\tilde{\psi}}{dc}\Big|^2\bigg],
\label{eq:unnormalizedqfi}
\end{eqnarray}
\begin{eqnarray}
F=\frac{4}{\mathcal{N}^2}\bigg[\!\int\!d\mathbf{x}\,\Big(\frac{d|\tilde{\psi}|}{dc}\Big)^{\!2}\!\!
-\Big(\frac{d\mathcal{N}}{dc}\Big)^{\!2}\bigg].
\label{eq:unnormalizedcfi}
\end{eqnarray}
\end{subequations}
If $\tilde{\psi}$ is a normalized wavefunction, i.e., $\mathcal{N}=1$,
then Eq.~\eqref{eq:qfi2} and Eq.~\eqref{eq:cfi2} are recovered.

The expansion of QFI with respect to the Bethe ansatz solution
under periodic boundary condition, i.e., Eq.~\eqref{eq:periodicLLsol}, is the following:
\begin{widetext}
\begin{eqnarray}
\mathfrak{F}_p=\frac{4}{\mathcal{N}^2}\Bigg[\,\,&&\sum_{P,Q}\bigg(\frac{dA^\ast(P)}{dc}\frac{dA(Q)}{dc}\,
\mathfrak{I}(k_P-k_Q)
+i\,\frac{dA^\ast(P)}{dc}A(Q)\sum_{l=1}^N\frac{dk_{Q_l}}{dc}\,\mathfrak{I}_l^1(k_P-k_Q)\nonumber\\
&&\qquad\quad\;\;\,-\,\,i\,A^\ast(P)\frac{dA(Q)}{dc}\sum_{l=1}^N\frac{dk_{P_l}}{dc}\,\mathfrak{I}_l^1(k_P-k_Q)
+A^\ast(P)A(Q)\!\!\!\sum_{m,n=1}^N\!\!\frac{dk_{P_m}}{dc}\frac{dk_{Q_n}}{dc}
\,\mathfrak{I}_{mn}^{11}(k_P-k_Q)\bigg)\nonumber\\
&&\,\,-\,\,\frac{1}{\mathcal{N}^2}\,\bigg|\sum_{P,Q}\Big(
A^\ast(P)\frac{dA(Q)}{dc}\,\mathfrak{I}(k_P-k_Q)
+i\,A^\ast(P)A(Q)\sum_{l=1}^N\frac{dk_{Q_l}}{dc}\,\mathfrak{I}_l^1(k_P-k_Q)\Big)\bigg|^2\Bigg],
\label{eq:periodicqfi}
\end{eqnarray}
\end{widetext}
where a notation
\begin{eqnarray}
\mathfrak{I}_{mn}^{\alpha\beta}(\lambda):=
\int_0^L\!\!\!dx_N\int_0^{x_N}\!\!\!\!\!\!dx_{N-1}\cdots
&&\int_0^{x_2}\!\!\!\!dx_1\,x_m^\alpha\,x_n^\beta\nonumber\\
&&e^{-i\sum_{j=1}^N\lambda_jx_j}\,,
\label{eq:multipleint}
\end{eqnarray}
is introduced to abbreviate similarly repetitive integrals
and the indices of $\mathfrak{I}$ are the integers satisfying $1\le m,n\le N$ and $\alpha,\beta\ge0$.
Also, $x_m^\alpha$ ($x_n^\beta$) means $x_m$ ($x_n$),
i.e., the position of $m$th ($n$th) particle, to the power of $\alpha$ ($\beta$)
and if the each power, i.e., $\alpha$ or $\beta$,  is zero,
then the corresponding subscript, i.e., $m$ or $n$, needs not be specified, thus it can be omitted.
The $\lambda$ in Eq.~\eqref{eq:multipleint} indicates $\{\lambda_1,\cdots,\lambda_N\}$
and likewise, $k_P$ and $k_Q$ are the short notations
for the permuted set of quasi-momenta, i.e., $(k_{P_1},\cdots,k_{P_N})$ and $(k_{Q_1},\cdots,k_{Q_N})$, repectively.
Hence all of $\mathfrak{I}(k_P-k_Q)$, $\mathfrak{I}_{l}^{1}(k_P-k_Q)$,
and $\mathfrak{I}_{mn}^{11}(k_P-k_Q)$
in Eq.~\eqref{eq:periodicqfi} are clearly defined.

The replacement of $\sum_{P,Q}\rightarrow\sum_{\epsilon,\delta}\pi_\epsilon\pi_\delta\sum_{P,Q}$,
$A(P)\rightarrow A(\epsilon,P)$, $A(Q)\rightarrow A(\delta,Q)$,
$k_{P_j}\rightarrow\epsilon_jk_{P_j}$, and $k_{Q_j}\rightarrow\delta_jk_{Q_j}$,
and modifying the definition of $\mathcal{N}$
from Eq.~\eqref{eq:periodicLLnorm} to Eq.~\eqref{eq:hardwallnorm}
lead to the corresponding QFI for the hard-wall boundary condition, i.e., $\mathfrak{F}_h$.
Refer to Eq.~\eqref{eq:periodicLLsol} and Eq.~\eqref{eq:hardwallLLsol} for the definitions
of the replaced symbols above.
The CFI in Eq.~\eqref{eq:unnormalizedcfi} can be similarly expanded
under each boundary condition.

In Eq.~\eqref{eq:periodicqfi},
it is obvious from the definitions that
$\mathcal{N}$, $A(P)$, $A(Q)$, and the multiple integral parts
all depend on the difference between any two quasi-momenta in $\{k_1,k_2,\cdots,k_N\}$,
not on the values of quasi-momenta themselves.
Furthermore, from the Bethe equations in Eq.~\eqref{eq:periodicBetheeq} under the periodic boundary condition,
we can derive the following analytical expression for the derivative of quasi-momenta
with respect to $c$:
\begin{equation}
\frac{dk_j}{dc}=\sum_{a=1}^N\big[\mathbf{H}^{-1}\big]_{ja}\sum_{l=1}^N\frac{2(k_a-k_l)}{c^2+(k_a-k_l)^2}\,,
\end{equation}
where the matrix $\mathbf{H}$ is from Eq.~\eqref{eq:periodicLLnorm}
and relies on the difference between any two quasi-momenta.
This shows that $dk_{P_j}/dc$ or $dk_{Q_j}/dc$ in Eq.~\eqref{eq:periodicqfi}
also depends on a set of $N-1$ elements $\{k_1-k_2,k_2-k_3,\cdots,k_{N-1}-k_N\}$,
not on $\{k_1,k_2,\cdots,k_N\}$.
The $dA(P)/dc$ and $dA(Q)/dc$ are functions of $\{k_1-k_2,k_2-k_3,\cdots,k_{N-1}-k_N\}$,
thus the QFI in Eq.~\eqref{eq:periodicqfi} is a function of $k_j-k_{j+1}$ for all $j$'s.
This implies that $\mathfrak{F}_p$ and $F_p$ are invariant
under the translation of whole quantum numbers:
$\{I_1,\cdots,I_N\}\rightarrow\{I_1+J,\cdots,I_N+J\}$.
The sum over permutations, i.e., $\sum_P\sum_Q$, makes $\mathfrak{F}_p$ and $F_p$ invariant
under the reverse of $\{k_1-k_2,k_2-k_3,\cdots,k_{N-1}-k_N\}$, too.
An example for this case is a transition from a state with quantum numbers $[-1,0,2]$
to a state with $[-1,1,2]$.

On the other hand, under the hard-wall boundary condition,
the derivative of Bethe equations in Eq.~\eqref{eq:hardwallBetheeq} gives
\begin{eqnarray}
\!\!\!\!\!\!\!\!\!\!\frac{dk_j}{dc}=&&\,\sum_{a=1}^N\big[\mathbf{H}^{-1}\big]_{ja}\nonumber\\
&&\,\,\,\,\times\sum_{l\neq a}\Big[\frac{k_a-k_l}{c^2+(k_a-k_l)^2}+\frac{k_a+k_l}{c^2+(k_a+k_l)^2}\Big],
\end{eqnarray}
where the matrix $\mathbf{H}$ is from Eq.~\eqref{eq:hardwallnorm}.
This means that the QFI and the CFI for the hard-wall geometry
depend on both of $\{k_j-k_{j+1}|j=1,2,\cdots,N-1\}$ and $\{k_j+k_{j+1}|j=1,2,\cdots,N-1\}$,
i.e., $\{k_j|j=1,2,\cdots,N\}$.
Thus there's no invariance of $\mathfrak{F}_h$ and $F_h$
with respect to the above operations on quantum numbers.

There are several difficulties in the calculation of QFI and CFI
with respect to the LL eigenstates.
One of them is that we encounter computing the multiple integrals
in the domain $D'$ as in Eq.~\eqref{eq:multipleint}.
These multiple integrals appear in the calculation of the norm, scalar product
\cite{Slavnov_1989,Chen_2020,Sotiriadis_ArXiv_2020},
and correlation function \cite{Zill_2016} with respect to the Bethe ansatz.
Using the simple calculative techniques specified in Ref.~\cite{Zill_2016}
\[
\int\!dx\,x^p\,e^{ikx}=-p!\,\Big(\frac{i}{k}\Big)^{p+1}\!\!e^{ikx}\sum_{s=0}^p\frac{(-ikx)^s}{s!}\,,
\]
we can reduce the computational time when compared 
to purely letting the integrals be implemented by Monte-Carlo simulation.
Another severe issue arises in the sums
which are inherent in the Bethe ansatz: $(\sum_P)^2$
or $(\sum_\epsilon\sum_P)^2$ for the periodic or hard-wall boundary conditions, respectively.
This results in a huge algorithmic complexity proportional to $\sim N!^2$ or $\sim 2^{2N}N!^2$
for each boundary condition.
To avoid this complexity, we restrict ourselves to a small number of particles
in the numerical calculation of Fisher information.

\section{\label{sec:analysis}Analysis of the Fisher information}

Since the CFI based on a wavefunction is expected to be attainable
by the absorption imaging method, which is later shown,
we calculate the QFI and the CFI for the interaction strength $c$
imprinted on the eigenfunctions of the LL Hamiltonian
to see how precisely the parameter can be determined
by implementing the single-shot measurement on the LL stationary states.
We investigate how under different boundary conditions
the Fisher information for $c$ behaves
according to the system variables, e.g., $c$, $L$, and $N$,
and the types of low-lying elementary excitations.
The Fisher information is numerically computed by Eq.~\eqref{eq:periodicqfi}
and its adjusted version for the hard-wall boundary condition.

\subsection{\label{subsec:analysis1}Ground states}

First we examine the case of the LL ground states
that have the least number of interacting particles, i.e., $N=2$.
Under the periodic boundary condition,
the quantum numbers for the ground state, i.e., $[I_1=-1/2,I_2=1/2]$,
lead to symmetrically distributed quasi-momenta, i.e., $k_2=-k_1$,
which make the wavefunction in Eq.~\eqref{eq:periodicLLsol} real-valued.
Under the hard-wall boundary condition,
the set of quantum numbers for the ground state is $[I_1=1,I_2=2]$
and results in the corresponding quasi-momenta satisfying $0<k_1<k_2$.
Due to the form of Bethe ansatz that has the wave components reflected from the hard-wall,
the wavefunction in Eq.~\eqref{eq:hardwallLLsol} is real-valued, too.
In conclusion, the QFI and the CFI are completely equal for the $N=2$ LL ground states.

\begin{figure}[t]
\includegraphics[width=.494\columnwidth]{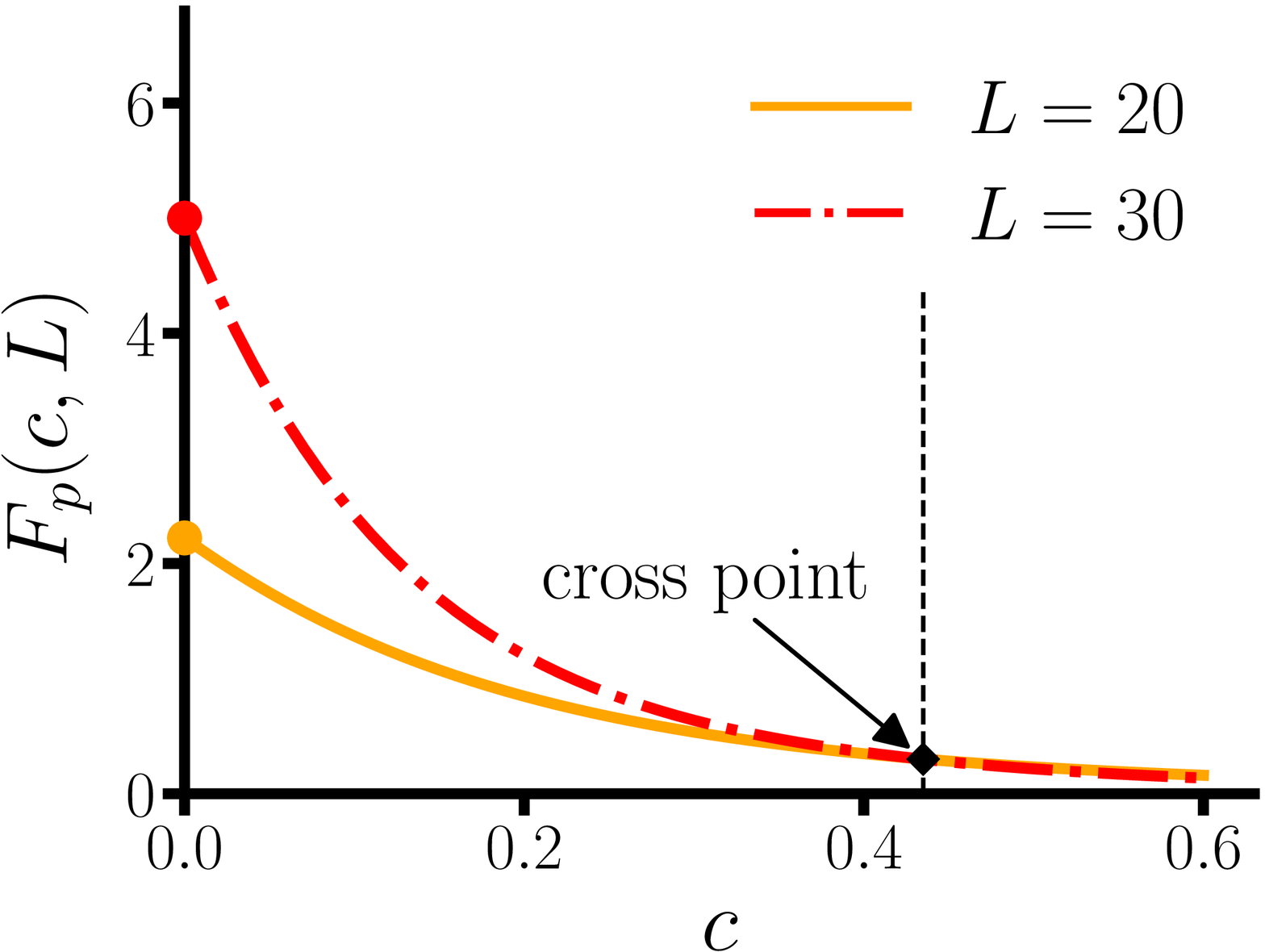}
\includegraphics[width=.494\columnwidth]{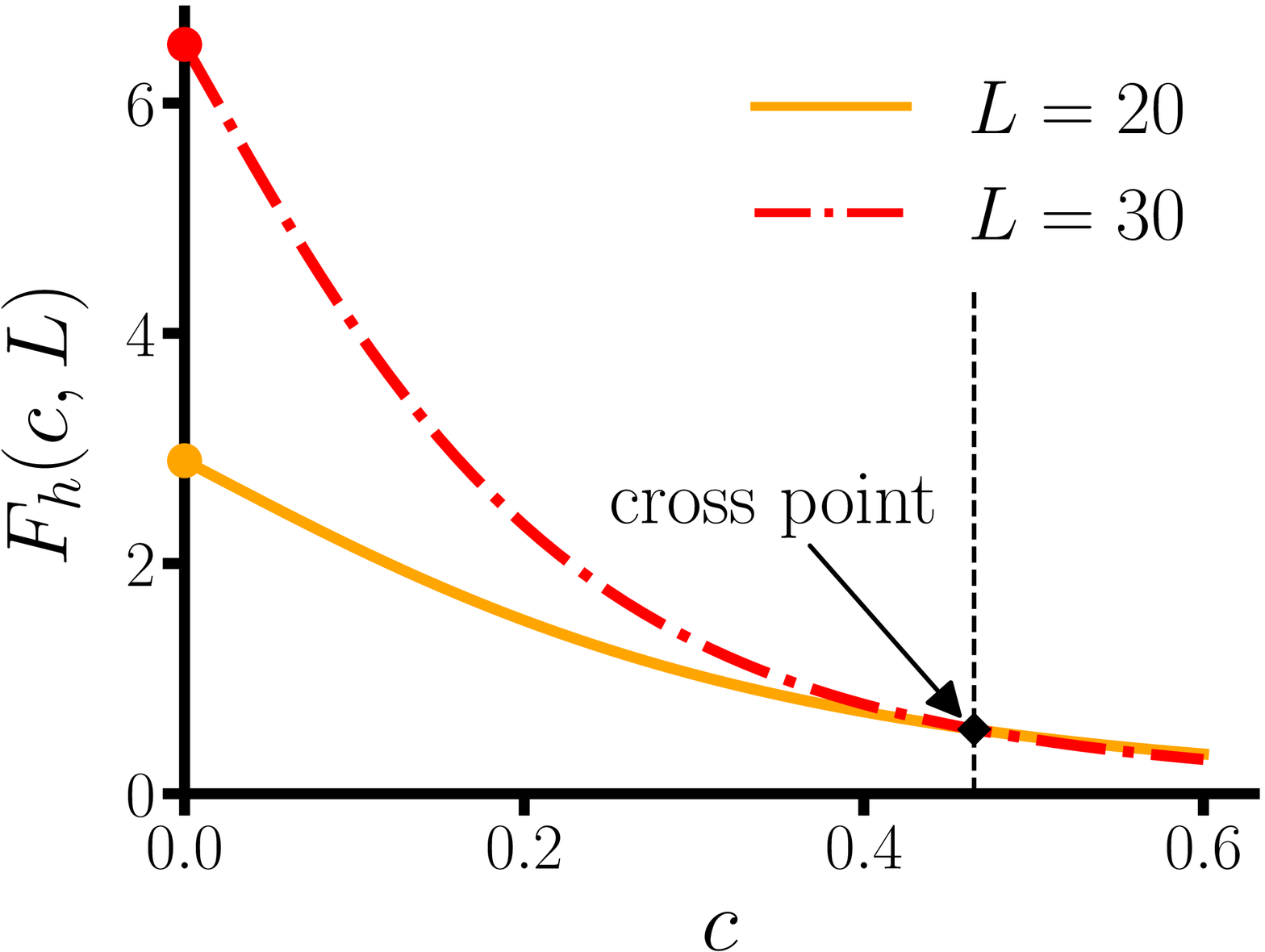}
\includegraphics[width=.494\columnwidth]{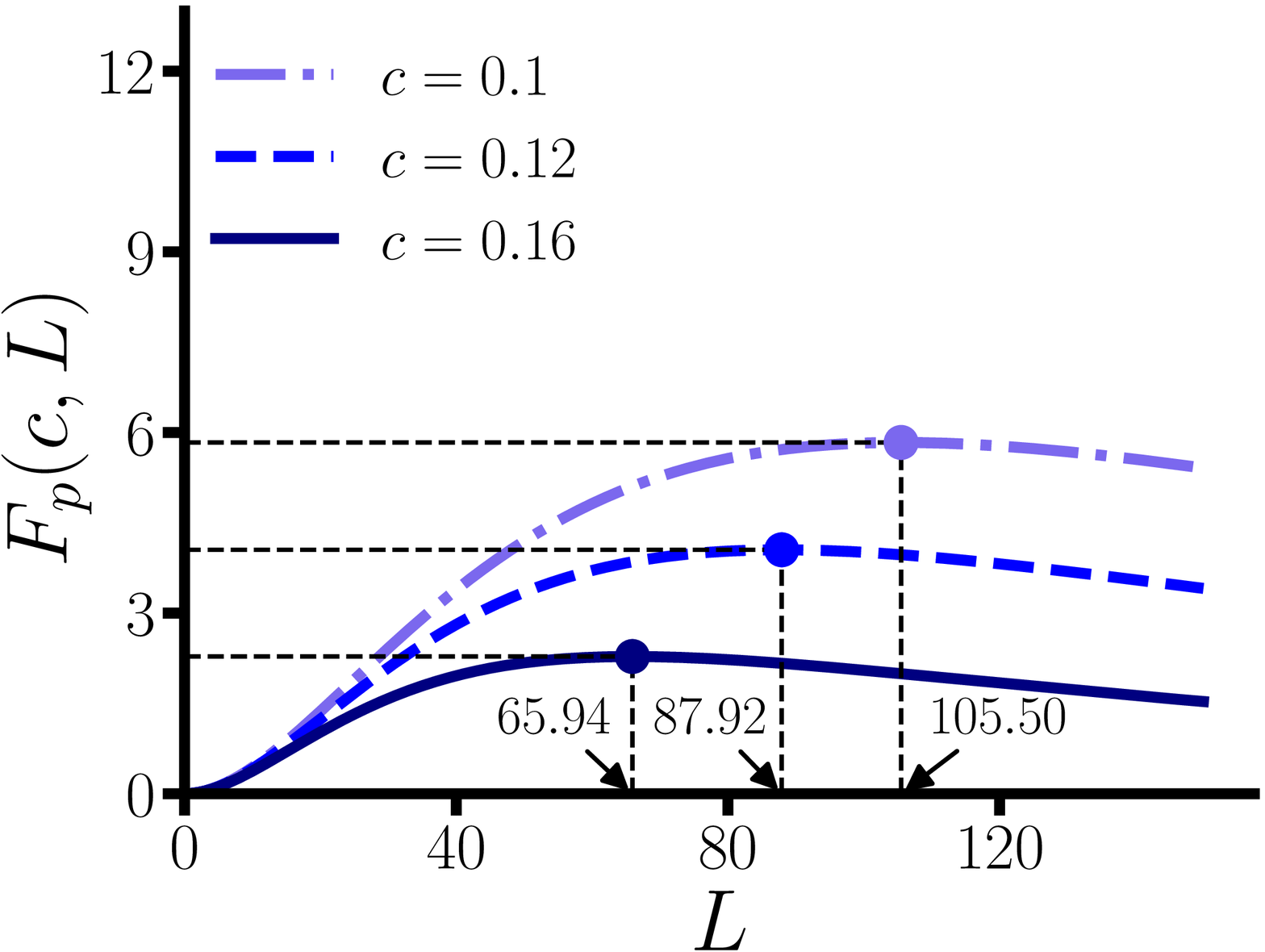}
\includegraphics[width=.494\columnwidth]{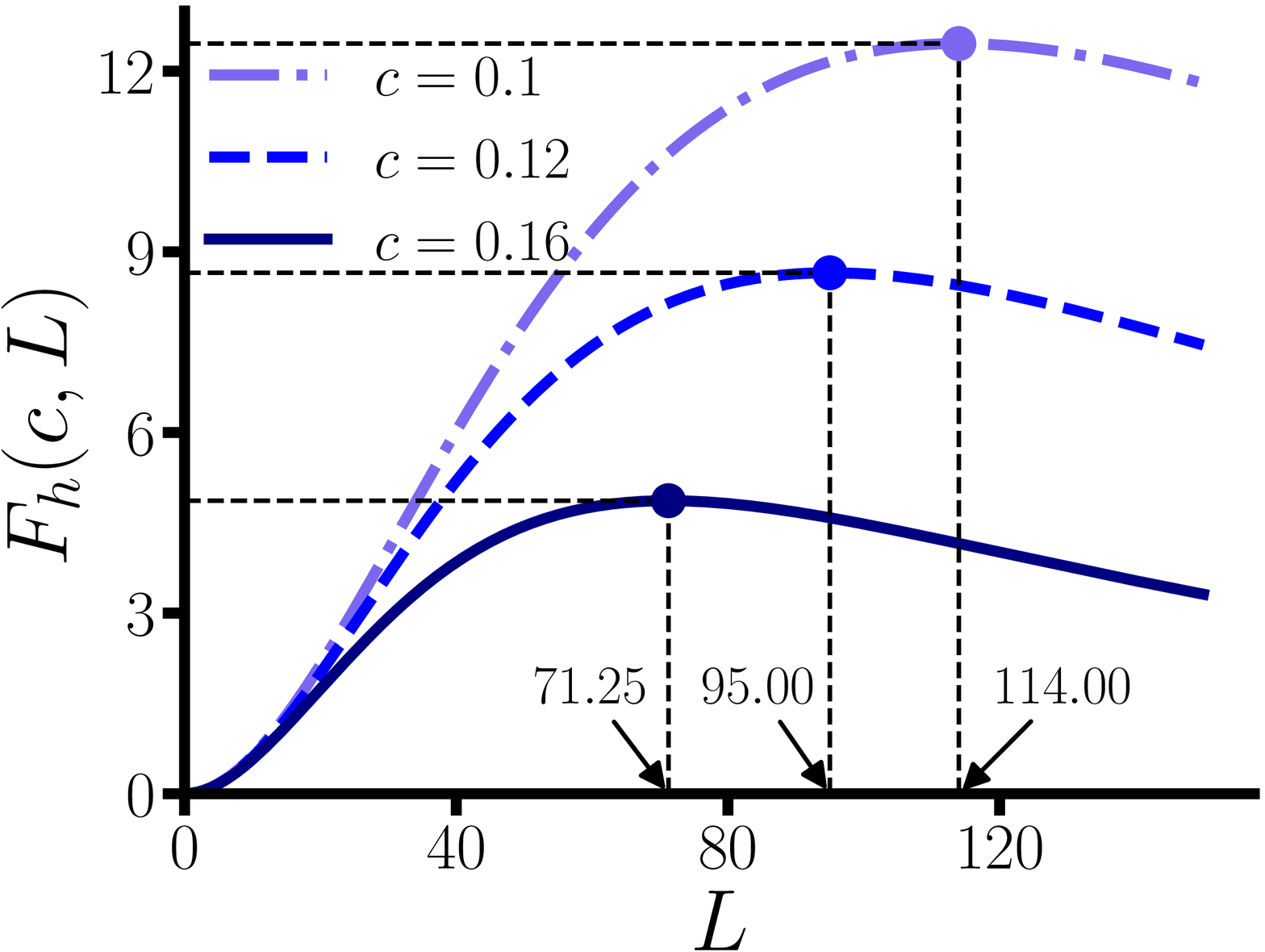}
\caption{\label{fig:result1}
The CFI for $c$ of the $N=2$ LL ground state
as a function of the system variables: $c$ and $L$.
In each row, the left plot is for the periodic boundary condition
and the right plot is for the hard-wall boundary condition.
For $N=2$, both ground-state wavefunctions are real-valued,
thus the CFI, i.e., $F(c,L)$, is equal to the QFI, i.e., $\mathfrak{F}(c,L)$.
In the upper row, the maximum value at $c=0$ is $L^2/180\simeq0.0056L^2$
for $F_p(c,L)$ and $L^2(-855+60\pi^2+4\pi^4)/(180\pi^4)\simeq0.0072L^2$
for $F_h(c,L)$. In the lower row, the maximum points appear
at $c\,L_\text{max}\simeq10.55$ for $F_p(c,L)$ and $c\,L_\text{max}\simeq11.40$ for $F_h(c,L)$.
Hence, for larger $c$, the optimal system size $L_\text{max}$ appears at a smaller value.
Also, note that the hard-wall geometry helps to enhance the Fisher information in excess of 
its ring-geometry counterpart.}
\end{figure}

The Fig.~\ref{fig:result1} shows how the CFI behaves
according to $c$ (upper row) or $L$ (lower row) in the case above.
In each row, the left plot shows the CFI for the periodic boundary condition, i.e., $F_p(c,L)$,
and the right plot shows the CFI for the hard-wall boundary condition, i.e., $F_h(c,L)$,
and they can be compared in parallel, with the same scales of axes.

In the upper row, we can see a rapid and monotonic decrement of CFI
as the interaction coupling $c$ increases, exhibiting a maximum at $c=0$.
Larger $L$ leads to higher CFI under the same value of $c$,
but this is reversed after some value of $c$. Also, note that $F_p(c,L)\leq F_h(c,L)$.
In the upper left plot,
the maximum value can be computed as $F_p(c=0,L)=L^2/180\simeq0.0056L^2$
by inserting the small-$c$ approximation
$k_j=q_j\sqrt{2c/L}+O(c)$, where the $q_j$ are the zeroes of Hermite polynomial of order $N$,
i.e., $H_N(q)$
\cite{Gaudin_2014}, into the $F_p(c,L)$ and taking the limit $c\rightarrow0$.
In the upper right plot,
$k_1=\pi/L-\sqrt{c/(2L)}+O(c)$ and $k_2=\pi/L+\sqrt{c/(2L)}+O(c)$
are inserted into $F_h(c,L)$, and then taking the limit $c\rightarrow0$ gives
$F_h(c=0,L)=L^2(-855+60\pi^2+4\pi^4)/(180\pi^4)\simeq0.0072L^2$.
When $c$ is extremely large, the quasi-momenta $k_j$'s are getting less relevant to the $c$
since $k_j\rightarrow2\pi I_j/L$ or $k_j\rightarrow\pi I_j/L$ up to the first dominant term
under the periodic or hard-wall boundary condition,
thus the distinguishability of the ground state with respect to $c$ decreases.

The lower row in Fig.~\ref{fig:result1} shows the variation of CFI
as the system size $L$ increases.
Both of $F_p(c,L)$ and $F_h(c,L)$ rise from $0$ at $L=0$,
reach their maxima at $L=L_\text{max}$, and finally decrease gradually.
As mentioned, the CFI is a quadratic function of $L$ when $c=0$
and monotonically increases as $L\rightarrow\infty$.
For any nonzero value of $c>0$, however, the region of the quadratic increase with $L$ of the CFI moves 
to a regime of smaller $L$, 
and the CFI begins to decrease after reaching a maximum at $L=L_\text{max}$,
thus it is confirmed that the interaction overall deteriorates the measurement precision.
The maximum of CFI gets lower as $c$ increases
and its position $L_\text{max}$ moves to the left accordingly.
We have numerically obtained this optimal value of $L$
by solving $dF_{p,h}(c,L)/dL|_{L=L_\text{max}}=0$:
$c\,L_\text{max}\simeq10.55$ for $F_p(c,L)$ and $c\,L_\text{max}\simeq11.40$ for $F_h(c,L)$.
In short, the optimal system size $L_\text{max}$ appears at a smaller value for a larger $c$.
Here, we also note that $F_p(c,L)\leq F_h(c,L)$,
where the maximal value of $F_h(c,L)$ is roughly two times larger than the one of $F_p(c,L)$
for the same $c$.

\begin{figure}[t]
\includegraphics[width=.494\columnwidth]{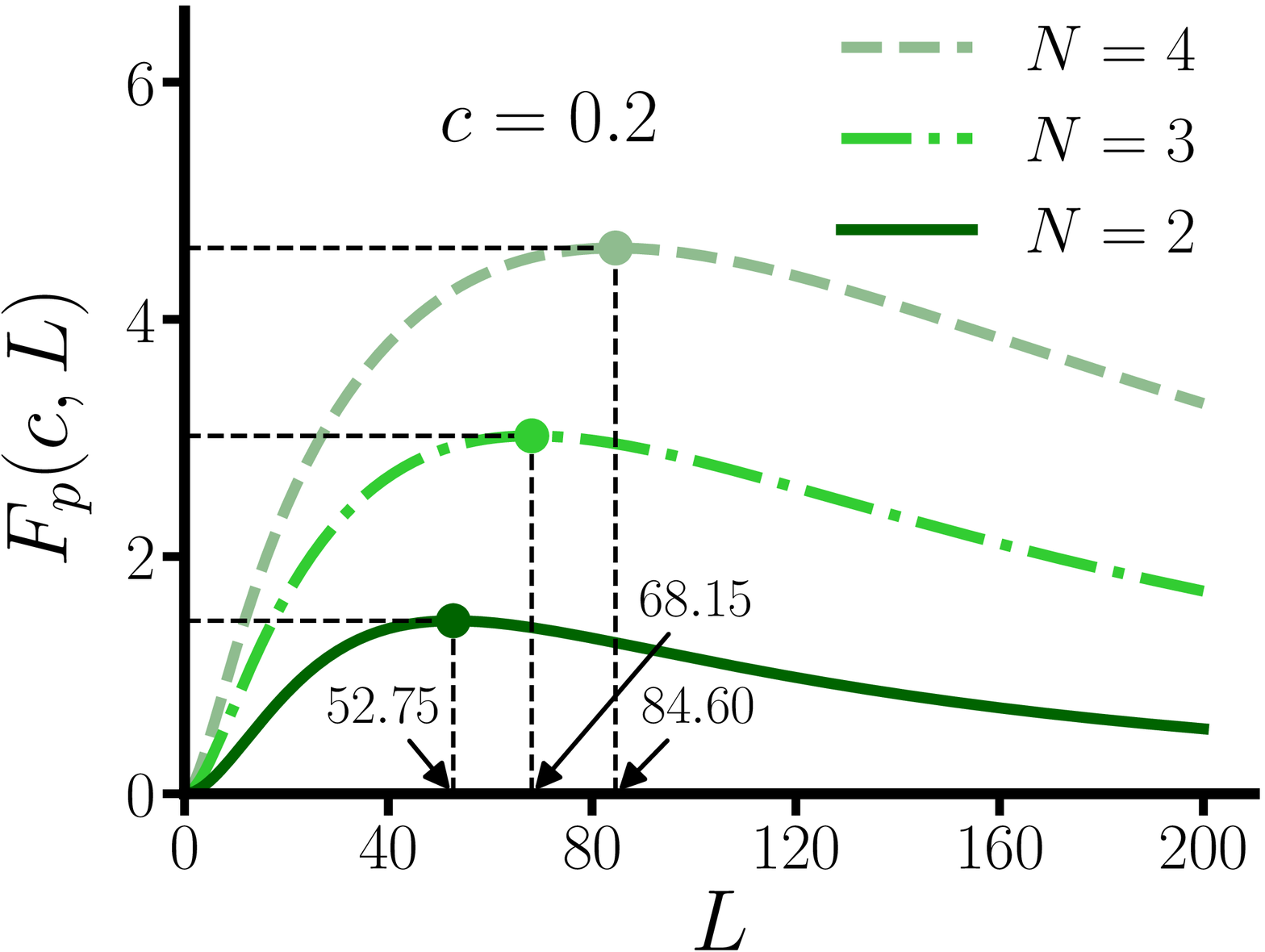}
\includegraphics[width=.494\columnwidth]{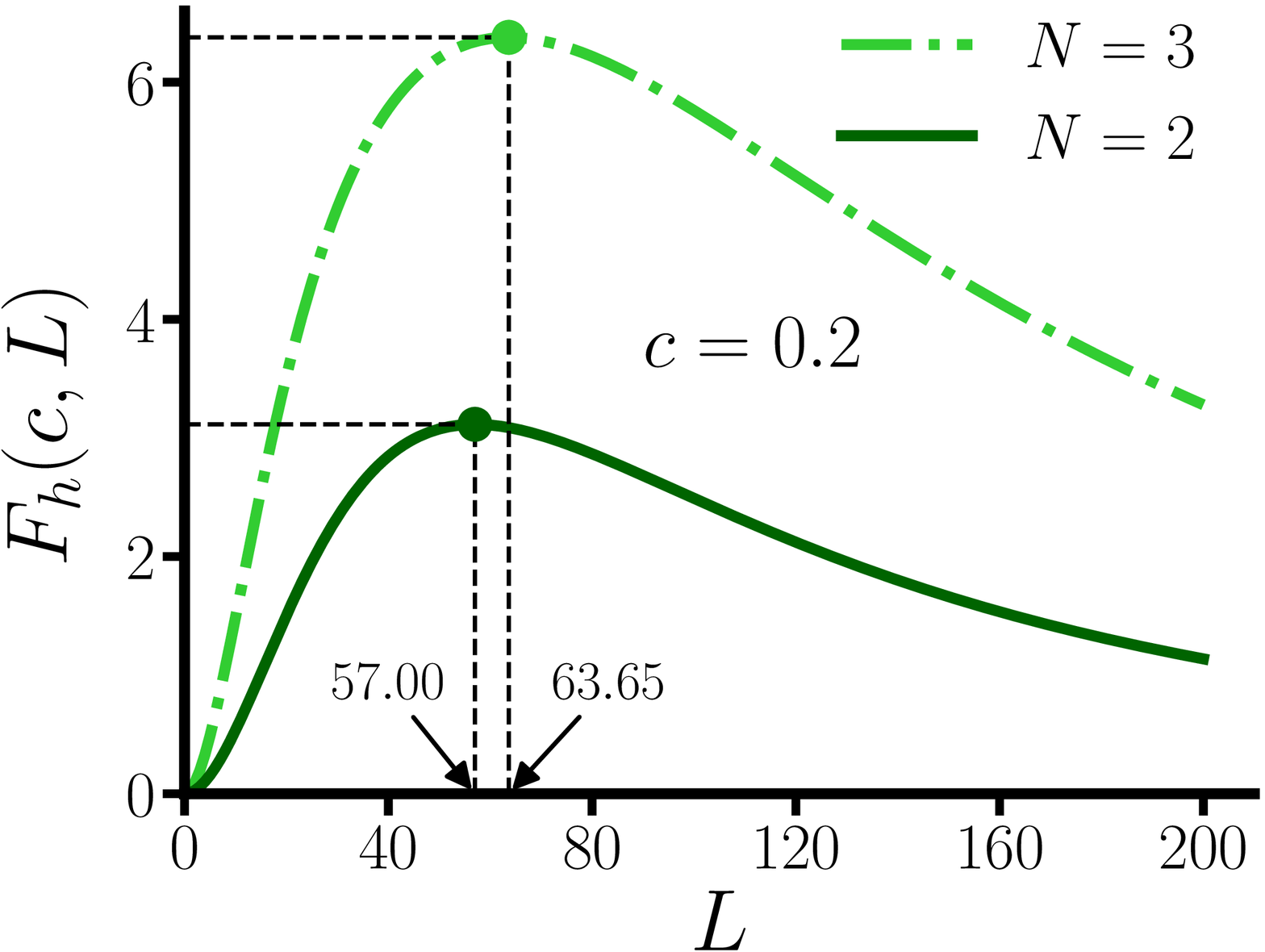}
\caption{\label{fig:result2}
The CFI for $c$ of the LL ground state as a function of $L$,
where $c=0.2$ and the particle number is slightly increased from $N=2$.
The left plot is for the periodic boundary condition,
in which the ground-state wavefunction is real-valued for all $N$.
The right plot is for the hard-wall boundary condition,
in which the ground-state wavefunction is real-valued for even $N$
or is purely imaginary-valued for odd $N$. In any case, the CFI is equal to the QFI.
As $N$ grows, the CFI shows a significant enhancement over the whole range of $L$
maintaining its behavioral pattern
and the $L_\text{max}$, at which the CFI is maximal, moves right:
$c\,L_\text{max}\simeq10.55,13.63,16.92$ for $F_p(c,L)$
and $c\,L_\text{max}\simeq11.40,12.73$ for $F_h(c,L)$.
For the same $N$, the system with hard-wall geometry
exhibits a much higher CFI.}
\end{figure}

In most cases, the Fisher information is dependent on a target parameter,
which leads to inhomogeneous estimation precision over the range of the target parameter.
In other words, the precision that would be obtained if the true $\theta$ were $\theta_1$ can be larger or smaller
than the one that would be attained if $\theta$ were a different value $\theta_2$.
An exception to this is the estimation of the mean of a Gaussian distribution,
where the Fisher information for the mean does not depend on the mean itself
and is given by the inverse of the variance.
Similarly, in Mach-Zehnder interferometry (MZI), 
one often tries to estimate the $\theta$ encoded by unitary operator, e.g., $e^{-i\theta\hat{\sigma}_y}$,
to a general quantum state $\rho_0=(I+\vec{s}_0\cdot\vec{\sigma})/2$,
where $\vec{s}_0\in\mathbf{R}^3$ with $|\vec{s}_0|\leq1$ and $\vec{\sigma}$ is the vector of Pauli matrices,
i.e., $\{\sigma_x,\sigma_y,\sigma_z\}$.
The QFI in this case depends only on $\vec{s}_0$, but not on $\theta$,
while the CFI for the measurement of population imbalance between two arms of interferometer,
which is the typical choice in MZI, is dependent on $\theta$ as well as the initial state parameter $\vec{s}_0$.
For pure states, there exists a class of optimal $\vec{s}_0$'s which lets the CFI saturate the QFI.
See Refs.~\cite{Barndorff-Nielsen_2000,Wasak_2016} for further details.
On the other hand, unlike the typical (linear) MZI above,
where the parameter to be estimated appears as a multiplicative factor
to the generator of unitary evolution,
the dependence of a Hamiltonian on the parameter can be more general \cite{Pang_2014}.
In this case, the QFI can rely on the target parameter itself.
In the present example of LL model,
the system size $L$ plays the role of initial state parameter
other than the target parameter $c$,
and the equality of QFI and CFI is guaranteed for the $N=2$ ground state.
Here, the QFI for $c$ has the dependence on $c$
as in the metrology with the general Hamiltonian above,
thus the optimal $L$, i.e., $L_\text{max}$, relies on the value of $c$,
which is quite different from the linear MZI case.
Also, our metrological protocol does not use any unitary operator to encode the target parameter
and that is why the QFI does not have a cyclical property with respect to $c$ \cite{Wasak_2016}.

In Fig.~\ref{fig:result2},
the particle number is raised up to $N=4$ in the case of $F_p(c,L)$ (left plot),
while up to $N=3$ in the case of $F_h(c,L)$ (right plot),
where $c=0.2$ is supposed in all plots.
The wavefunction in Eq.~\eqref{eq:periodicLLsol} is real-valued for any $N$
if the quasi-momenta, i.e., $\{k_1,k_2,\cdots,k_N\}$, are symmetrically distributed around $0$
and the ground state satisfies this condition.
On the other hand, the wavefunction in Eq.~\eqref{eq:hardwallLLsol} is real-valued for even $N$
and purely imaginary-valued for odd $N$.
Thus, as far as the ground-states are concerned, the CFI and the QFI are equal for any $N$.
As $N$ increases, the CFI increases over the whole range of $L$
and the $L_\text{max}$ increases accordingly:
$c\,L_\text{max}\simeq10.55,13.63,16.92$ for $F_p(c,L)$
and $c\,L_\text{max}\simeq11.40,12.73$ for $F_h(c,L)$.
Also, the advantage of hard-wall geometry for the CFI is clear from the Fig.~\ref{fig:result2},
when keeping $N$ fixed.

\subsection{\label{subsec:analysis2}Excited states}

\begin{figure}[t]
\includegraphics[width=.494\columnwidth]{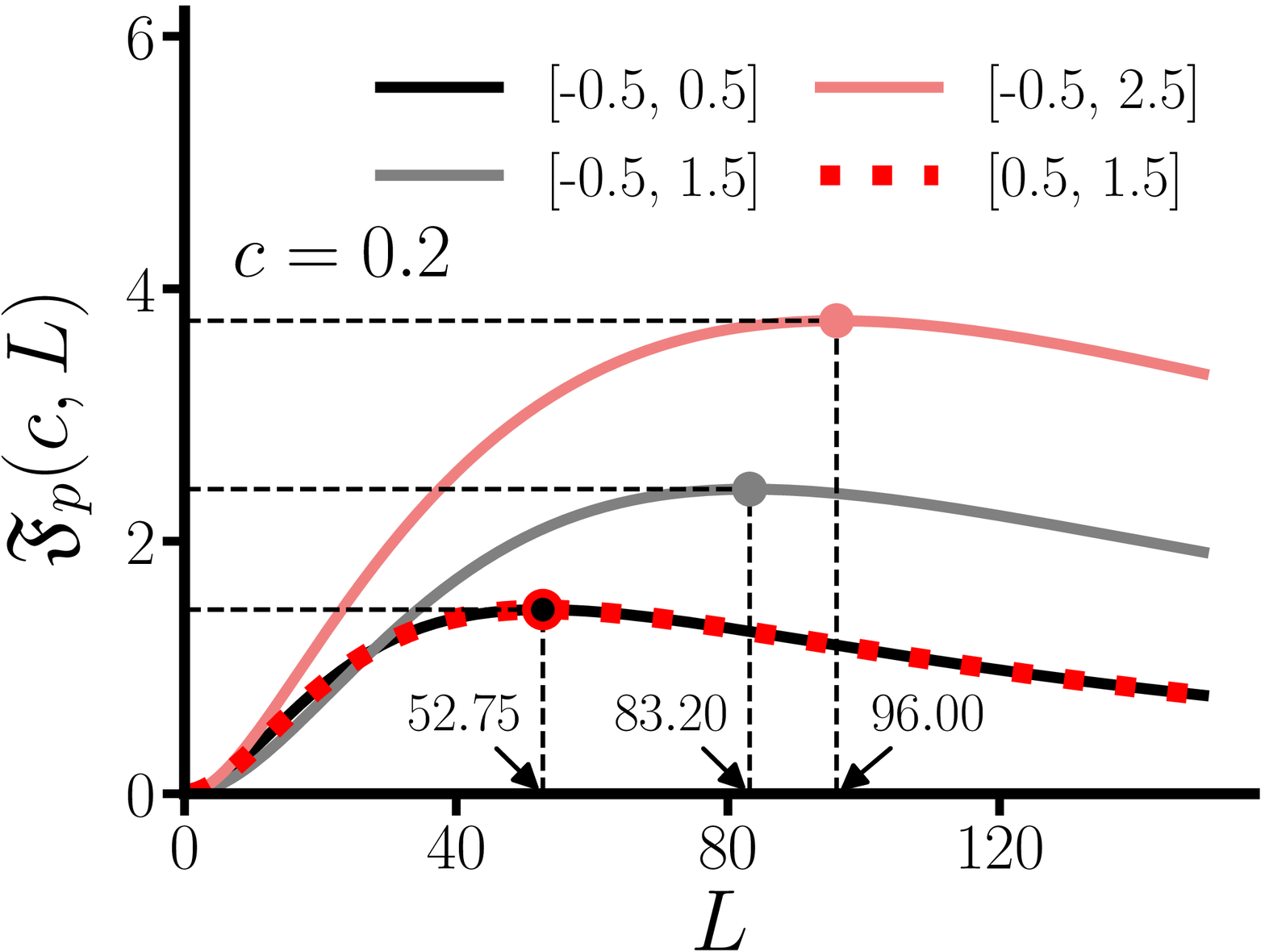}
\includegraphics[width=.494\columnwidth]{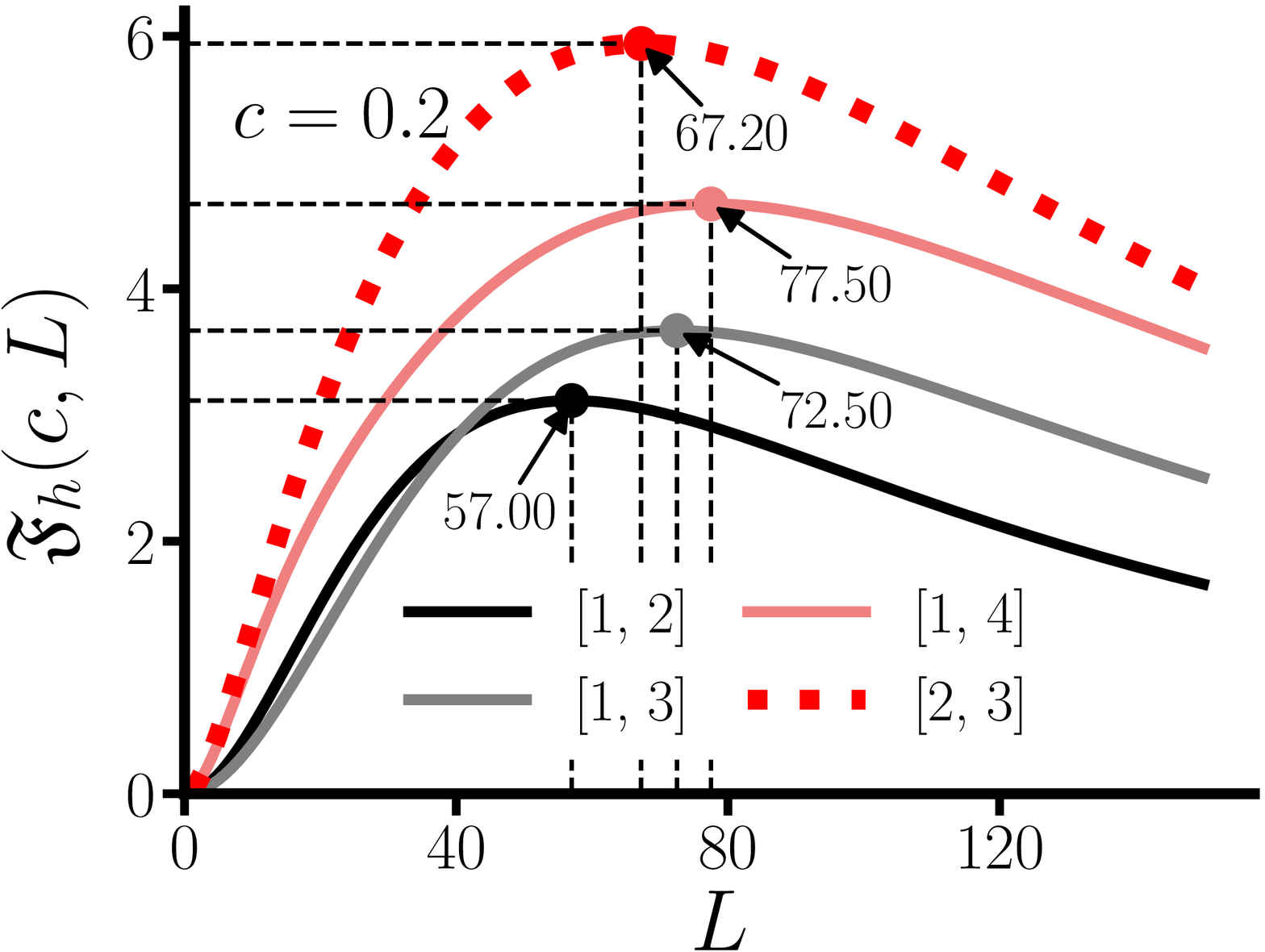}
\includegraphics[width=.494\columnwidth]{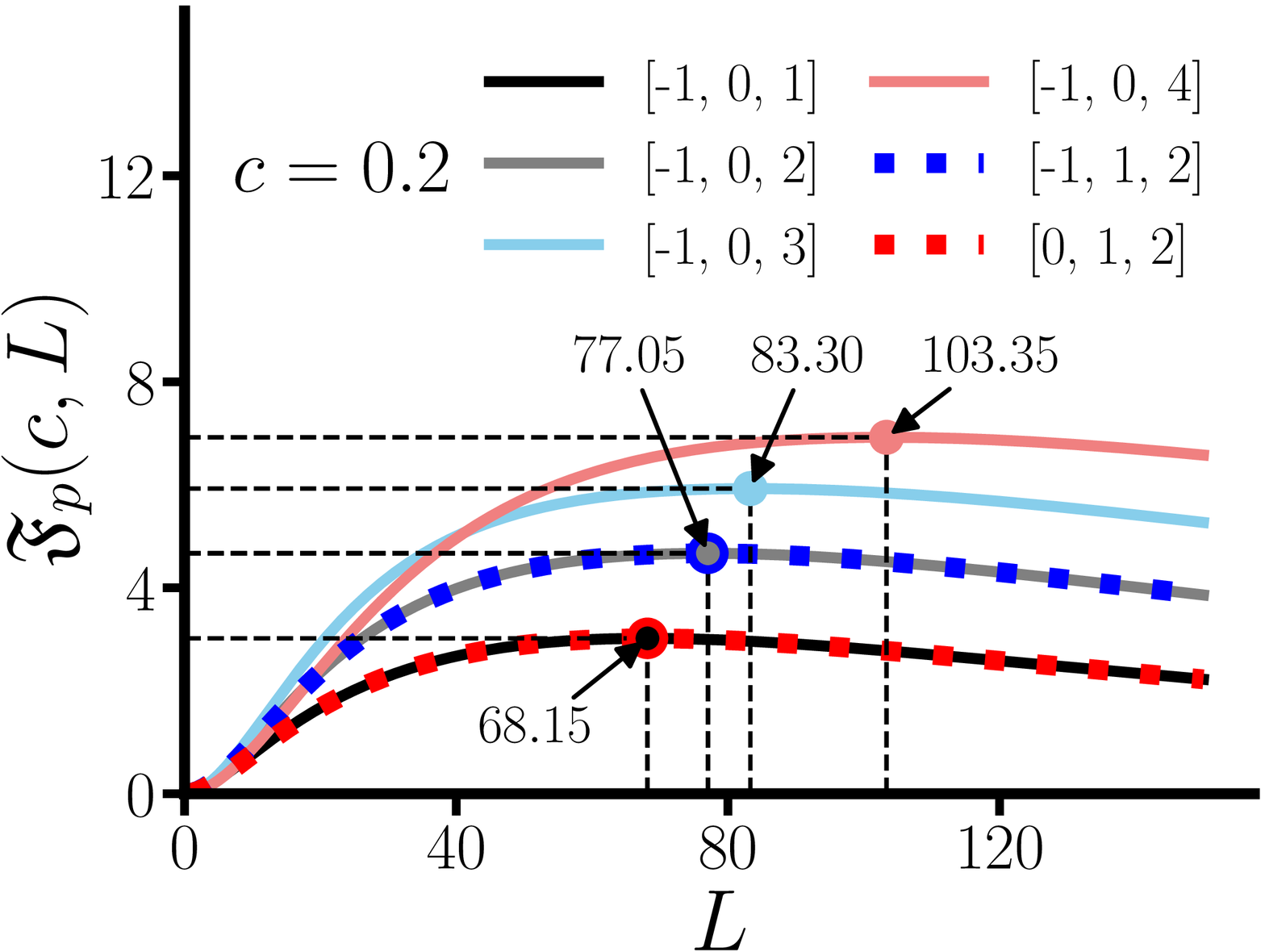}
\includegraphics[width=.494\columnwidth]{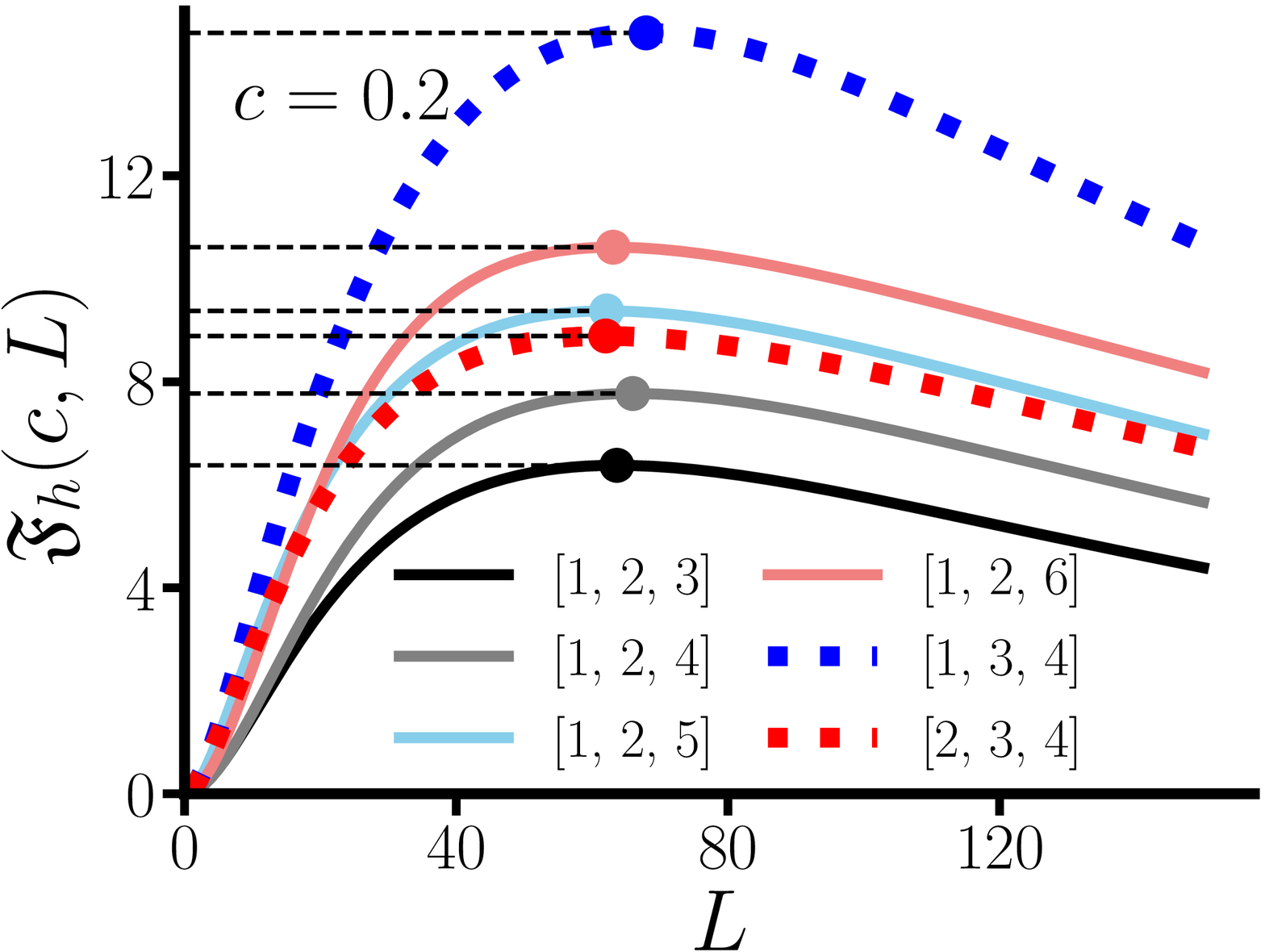}
\caption{\label{fig:result3}
The QFI for $c$ of the LL ground or excited state
as a function of $L$, where $c=0.2$.
The upper row is for $N=2$ and the lower row is for $N=3$.
In each row, the left plot is for the periodic boundary condition
and the right plot is for the hard-wall boundary condition.
Upper left) The type-I excited states give higher QFI as the energy increases,
but the type-II excited state, i.e., $[0.5,1.5]$ (red dotted),
gets back to the level of ground state, i.e., $[-0.5,0.5]$ (black solid).
Upper right) The system with hard-wall geometry similarly exhibits higher QFI
as it gets more type-I excited, but now the type-II excited state, i.e., $[2,3]$ (red dotted),
shows much higher QFI than its type-I counterpart (lightred solid).
Lower left) The QFI of any type-II excited state is always smaller
than that of its type-I counterpart and repeats the QFI of ground or type-I excited state.
Lower right) The type-II excited state with $[1,3,4]$ (blue dotted) gives the best QFI,
but another one with $[2,3,4]$ (red dotted) gives lower QFI than its type-I counterpart (lightred solid).
The values of $L_\text{max}$ are 63.65 (black solid), 66.00 (gray solid),
62.15 (lightblue solid), 63.10 (lightred solid), 67.95 (blue dotted), and 62.00 (red dotted).}
\end{figure}

For low-lying excitations,
there are two branches of elementary excitations in the LL model \cite{Lieb_1963}: 
the type-I and the type-II varieties,
which are distinguished by the energy-momentum dispersion relation.
For a given total momentum, type-I excited state is the state of maximal energy
and type-II excited state is the state of minimal energy.
Those are also identified by two different types of the transition of quantum numbers
from $I_{j=1,\cdots,N}=-(N-1)/2+j-1$ or $I_{j=1,\cdots,N}=j$,
which represents the distribution of quantum numbers of the ground state
under periodic or hard-wall boundary condition.
For the periodic boundary condition,
$I_{j=1,\cdots,N-1}=-(N-1)/2+j-1$ and $I_N=(N-1)/2+q$ with an integer $q\geq 1$
represent the type-I excited states.
The type-II excited states are represented
by $I_{j<q}=-(N-1)/2+j-1$ and $I_{j\geq q}=-(N-1)/2+j$ with an integer $1\leq q\leq N-1$.
The type-II excitation with $q=1$ is in fact classified
as the third type excitation called Umklapp excitation,
but for our discussion this can be understood as an extension of the type-II excitation.
For the hard-wall boundary condition,
the total momentum is not defined since it is not a conserved quantity.
However, we can find an appropriate alternative resembling the total momentum
\cite{Syrwid_2021}: $P=(\pi/L)\sum_{j=1}^N(I_j-j+1)$,
with which the energy-momentum dispersion relation can be established.
Then the type-I excitation is defined by
$I_{j=1,\cdots,N-1}=j$ and $I_N=N+q$ with an integer $q\geq1$.
The type-II excitation is expressed as
$I_{j<q}=j$ and $I_{j\geq q}=j+1$ with an integer $1\leq q\leq N-1$.

The Fig.~\ref{fig:result3} shows the QFI for $c$ of the LL ground or excited state
as a function of $L$, where $c=0.2$.
The upper row is for $N=2$ and the lower row is for $N=3$.
In each row, the left plot is for the periodic boundary condition
and the right plot for the hard-wall boundary condition.
The solid lines exhibit ground and type-I excited states,
while dotted lines are for the type-II excited states.
A pair of type-I and type-II excited states that have the same total momentum
is similarly colored, i.e., lightred $\leftrightarrow$ red and lightblue $\leftrightarrow$ blue.
Each state is denoted by a set of quantum numbers.
As explained, $\mathfrak{F}_p(c,L)$ is equal to $F_p(c,L)$
when the quantum numbers are symmetric around $0$,
because the wavefunction is real-valued then. 
When $N=2$, however, $\mathfrak{F}_p(c,L)=F_p(c,L)$ holds
for any set of quantum numbers
due to the invariance of $\mathfrak{F}_p(c,L)$ and $F_p(c,L)$
with respect to the translation of quantum numbers.
For example, $[I_1=0.5,I_2=2.5]$ can be translated into $[I_1=-1,I_2=1]$
while keeping $\mathfrak{F}_p(c,L)$ and $F_p(c,L)$ constant,
and the wavefunction for $[I_1=-1,I_2=1]$, which is artificially made symmetric around $0$,
is clearly real-valued.
Thus it is concluded that $\mathfrak{F}_p(c,L)=F_p(c,L)$ even for $[I_1=0.5,I_2=2.5]$.
When $N>2$, there are many states whose quantum numbers cannot be made symmetric around $0$,
hence $\mathfrak{F}_p(c,L)>F_p(c,L)$ by Eq.~\eqref{eq:qfi4}.
However, in current regimes of parameters, the difference is confirmed to be minor
and we ignore it in the following discussion.
On the other hand, $\mathfrak{F}_h(c,L)=F_h(c,L)$ is guaranteed for all $N$
since the wavefunction in Eq.~\eqref{eq:hardwallLLsol} is always
either of real-valued or pure imaginary-valued, as discussed in Sec.~\ref{subsec:fisherinfo}.
We can see in Fig.~\ref{fig:result3} how the QFI for $c$ behaves
as the state shifts from the ground state to the different excited states.

In the upper left plot,
as the energy increases along the type-I excitations,
the $\mathfrak{F}_p(c,L)$ shows a higher maximum with a larger $L_\text{max}$.
When the state is excited to $[I_1=-0.5,I_2=1.5]$ (gray solid),
the $\mathfrak{F}_p(c,L)$ slightly decreases in the lower range of $L$,
but in the upper range it shows a better improvement.
The lightred solid line is a type-I excited state with $[I_1=-0.5,I_2=2.5]$
and the red dotted line is its type-II counterpart, i.e., $[I_1=0.5,I_2=1.5]$, with the same total momentum:
$P=(2\pi/L)\sum_{j=1}^NI_j=4\pi/L$.
Here, we see that the type-II excitation worsens the QFI compared to its type-I counterpart.
Because of the dependence of $\mathfrak{F}_p(c,L)$ on $k_1-k_2$, or, in other words, $I_1-I_2$,
the $\mathfrak{F}_p(c,L)$ for $[I_1=1.5,I_2=2.5]$ (red dotted)
coincides with the one for $[I_1=-0.5,I_2=0.5]$ (black solid).
On the other hand, in the upper right plot,
the $\mathfrak{F}_h(c,L)$ for a type-II excited state with $[I_1=2,I_2=3]$ (red dotted),
exhibits a significant improvement with higher maximum but less $L_\text{max}$
compared to its type-I counterpart, i.e., $[I_1=1,I_2=4]$ (lightred solid).
Contrary to the periodic boundary condition,
the type-II excitation contributes better to enhance the precision limit than its type-I counterpart does
under the hard-wall boundary condition.

The lower row in Fig.~\ref{fig:result3} produces similar results for $N=3$ just as the upper row.
In the lower left plot,
the maximum of $\mathfrak{F}_p(c,L)$ and the $L_\text{max}$ increase
as the state is getting type-I excited.
However, all of type-II excited states have lower QFIs than those of their type-I counterparts.
For a given total momentum of $P=4\pi/L$,
the QFI for a type-II excited state with $[I_1=-1,I_2=1,I_3=2]$ (blue dotted) is less than the one
for a type-I excited state with $[I_1=-1,I_2=0,I_3=3]$ (lightblue solid)
and is equal to the QFI for a type-I excited state with $[I_1=-1,I_2=0,I_3=2]$ (gray solid),
which is due to the invariance of $\mathfrak{F}_p(c,L)$ with respect to the reverse of the set $[I_1-I_2,I_2-I_3]$
as explained in Sec.~\ref{subsec:}.
Also, for the total momentum of $P=6\pi/L$,
the QFI for a type-II excited state with $[I_1=0,I_2=1,I_3=2]$ (red dotted)
is smaller than the one for a type-I excited state with $[I_1=-1,I_2=0,I_3=4]$ (lightred solid)
and is equal to the QFI for the ground state, i.e., $[I_1=-1,I_2=0,I_3=1]$ (black solid),
which is attributed to the invariance of $\mathfrak{F}_p(c,L)$
with respect to the translation of quantum numbers.
The lower right plot shows
that the QFI for a type-II excited state does not always surpass the QFI of its type-I counterpart.
The type-II excited state with $[I_1=1,I_2=3,I_3=4]$ (blue dotted) reaches the highest QFI,
but the type-II excited state with $[I_1=2,I_2=3,I_3=4]$ (red dotted) exhibits less QFI than
its type-I counterpart, i.e., $[I_1=1,I_2=2,I_3=6]$ (lightred solid).
The $\mathfrak{F}_h(c,L)$ depends on each value of quasi-momenta,
not on the difference between any two quasi-momenta,
thus, unlike the periodic boundary condition, identical QFIs do not exist.

We see the difference between two trap geometries 
in mainly that the hard-wall geometry is more favorable to the estimation of $c$
and that a type-II excitation leads to a further enhancement of the precision limit
under the hard-wall boundary condition.

\subsection{\label{subsec:analysis3}Fisher information for absorption imaging}

\begin{figure}[t]
\includegraphics[width=.494\columnwidth]{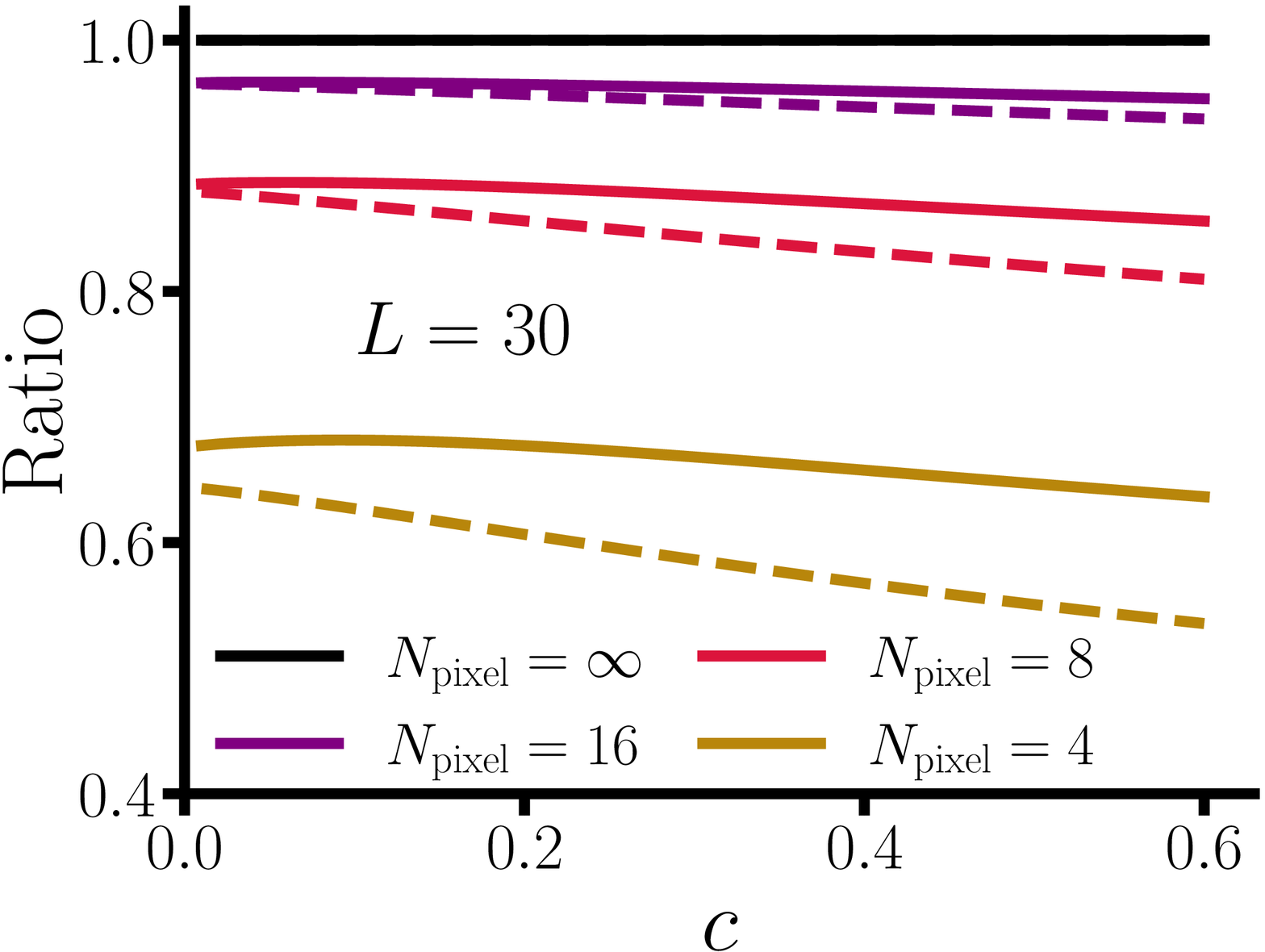}
\includegraphics[width=.494\columnwidth]{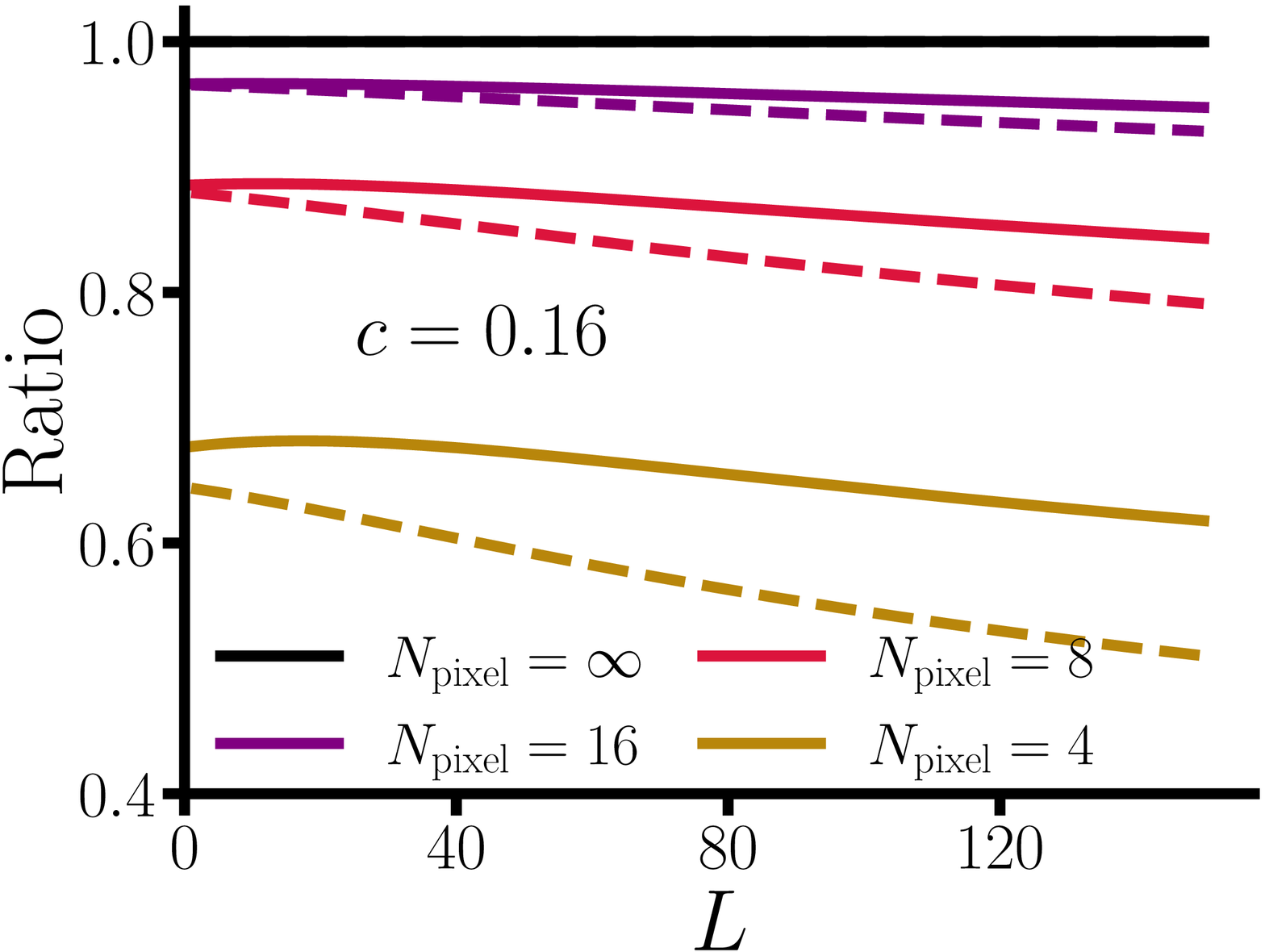}
\caption{\label{fig:result4}
The ratio between the CFI for absorption imaging with finite $N_\text{pixel}$
and the $F(c,L)$, i.e., the CFI when $N_\text{pixel}=\infty$, in the space of system variables: $c$ and $L$.
As the number of pixels in the absorption imaging increases,
the ratio is getting closer to 1 (black solid),
thus the imaging method results in the precision provided by the $F(c,L)$.
For each color,
the dashed line is for the periodic boundary condition
and the solid line is for the hard-wall boundary condition.
When estimating a parameter from the absorption imaging method,
the hard-wall geometry gives a larger CFI than that of the ring geometry
and also, the former deteriorates much less as $c$ or $L$ grows.}
\end{figure}

As explained, the absorption imaging is commonly implemented
to investigate the characteristics of ultracold atomic systems.
This method involves irradiating light to cold atomic gas from above,
extracting the optical density captured by the CCD camera below the gas,
and converting the optical density to the column density of atoms
located above each pixel of the camera.
In Sec.~\ref{subsec:measure}, this process has been modeled as dividing the finite one-dimensional space
into $N_\text{pixel}$ pixels and counting the number of atoms in each pixel bin.
We can calculate the CFI for this absorption imaging method
and compare it to $F_p(c,L)$ or $F_h(c,L)$,
which corresponds to the case when $N_\text{pixel}\rightarrow\infty$.
The Fig.~\ref{fig:result4} shows the convergence of CFI for the absorption imaging
as the number of pixels increases.
The y-axis represents the ratio between the CFI for the absorption imaging
and $F(c,L)$. The $F(c,L)$ is calculated using $N=2$ ground state in the LL model.
The left plot is when $L$ is fixed and the right plot is when $c$ is fixed.
The solid lines indicate the hard-wall geometry
and the dashed lines represent the ring geometry.
The CFI for the absorption imaging is higher in the hard-wall geometry
than in the ring geometry,
while both go to the ideal limit, i.e., $F(c,L)$, as $N_\text{pixel}$ increases.
When $c$ or $L$ increases, the saturation of the CFI for the absorption imaging
to the $F(c,L)$ tends to deteriorate, reflecting the Tonks-Girardeau behavior of gas
in the $\gamma\rightarrow\infty$ limit.
However, the hard-wall geometry shows less deterioration than the ring geometry
in this region.

\section{\label{sec:conclusion}Summary and Conclusion}

We have investigated the Fisher information of the interaction strength in the LL model
by calculating it with respect to the eigenstates of the LL Hamiltonian.
First we briefly reviewed the Bethe ansatz solutions of the LL model
under two different boundary conditions: periodic and hard-wall boundary conditions.
Then we discussed the QFI and the CFI in terms of the wavefunction
and the condition for the CFI to saturate the QFI.
We numerically calculated the Fisher information of interaction strength
using the LL eigenfunctions, i.e., Bethe ansatz,
and described its behavior in the space of system variables such as interaction strength and system size.
We confirmed the negative effect of interaction on the measurement precision
and, in particular, found an optimal value of system size for the estimation of interaction strength
and the advantage of the hard-wall geometry over the ring geometry.
All of these results were reproduced
for varying number of particles and with the LL eigenstates of different energies.
For the low-lying excitations,
we explained the different aspects of type-I and type-II excited states
appearing in the functional profile of Fisher information
based on its invariance
with respect to some symmetry operations on a set of quantum numbers.
Finally, we showed that the CFI supported by the absorption imaging method converges to the wavefunction-based CFI
as the pixel size in the model of absorption imaging improves.
This demonstrates
that the precision achievable by the ideal position measurement on $N$ particles can be attained
by elaborating the resolution of absorption imaging.

From the viewpoint of quantum metrology,
our analysis is selecting a different metrological framework from the typical protocol,
where a Hamiltonian parameter is imprinted by the unitary process on a quantum state,
which is initially irrelevant to the parameter to be estimated.
Here we rather try to determine the attainable precision of estimating a Hamiltonian parameter
using the eigenstates of that Hamiltonian
which already contain the information of the target parameter.
Within this framework,
instead of improving the $N$-scaling of Fisher information,
we have focused on its behavior in the parameter space
and the precision limit that can be obtained by a realistic measurement,
e.g., absorption imaging, 
which is the representative measurement to explore the ultracold atomic systems.
It is expected that these results can be utilized in the technique of adaptive measurement, 
to enhance the estimation precision of the interaction strength in ultracold bosonic gases.

There have been numerous investigations of the LL model 
under two extreme limits of the Lieb's parameter $\gamma$:
weakly and strongly interacting regimes,
which are defined by $\gamma\rightarrow0$ and $\gamma\rightarrow\infty$, respectively.
However, we confirmed that the optimal value of the resource parameter, i.e., system size,
that maximizes the Fisher information of the target parameter, i.e., interaction strength,
appears in the intermediate regime: $\gamma\sim O(1)$.
Hence the present results for the Fisher information,
highlighting the distinguishability of the Bethe ansatz solutions for varying $\gamma$, 
reveal that the intermediate regime of the LL model is in fact the most intriguing one, 
as demonstrated by its metrological usefulness.

\begin{acknowledgments}
This work has been supported by the National Research Foundation of Korea under 
Grants No.~2017R1A2A2A05001422 and No.~2020R1A2C2008103.
\end{acknowledgments}

\begin{widetext}
\appendix*

\section{Bosonic gas in a box trap}

The exact solution for $N$ bosons in a box of length $L$ is summarized here.
The system is formulated by $\hat{H}\,\Psi=E\,\Psi$ with the following Hamiltonian and boundary conditions:
\begin{subequations}
\begin{equation}
\hat{H}=-\sum_{j=1}^N\frac{\partial^2}{\partial x_j^2}+2\,c\sum_{j<l}\delta(x_j-x_l)\,,
\end{equation}
\begin{equation}
\Psi(x_1=0,x_2,\cdots,x_N)=\Psi(x_1,\cdots,x_{N-1},x_N=L)=0\,,
\label{eq:hardwallbc}
\end{equation}
\end{subequations}
where $c>0$ and $\hbar=2m=1$ for the units.
The $\Psi$ is symmetric under the exchange of arbitrary two spatial coordinates,
thus it is fully defined in the restricted real space $[0,L]^N$
once defined in the domain $D:0\le x_1\le x_2\le\cdots\le x_N\le L$.

The (unnormalized) solution for the domain $D$ is given by the Bethe ansatz
of the following form \cite{Gaudin_1971,Batchelor_2005,Gaudin_2014}:
\begin{eqnarray}
\label{eq:hardwallsol}
\tilde{\psi}(x_1,\cdots,x_N)=\!\!\sum_{\epsilon_1,\cdots,\epsilon_N}&&\!\sum_P\epsilon_1\epsilon_2\cdots\epsilon_N
A(\epsilon_1k_{P_1},\cdots,\epsilon_Nk_{P_N})\,e^{i\sum_{j=1}^N\epsilon_jk_{P_j}x_j}\,,\\
A(\epsilon_1k_{P_1},\cdots,\epsilon_Nk_{P_N})\;&&=\prod_{j<l}\Big(1-\frac{i\,c}{\epsilon_jk_{P_j}+\epsilon_lk_{P_l}}\Big)
\Big(1+\frac{i\,c}{\epsilon_jk_{P_j}-\epsilon_lk_{P_l}}\Big)\,,\nonumber
\end{eqnarray}
in which $\epsilon_j$ is summed over $\pm1$ and $P$ runs over all permutations of $[1,2,\cdots,N]$.
The sign of $k_j$ is absorbed into $\epsilon_j$ to make $k_j>0$ for all $j$'s
and also $k_1<k_2<\cdots<k_N$ is assumed without loss of generality due to the sum over $P$.
Now one can find the corresponding energy $E=\sum_{j=1}^Nk_j^2$
when $x_j\neq x_{j+1}\,(\text{i.e.}\,x_1<x_2<\cdots<x_N)$
and the rule of exchanging neighboring arguments of $A$:
\begin{equation}
A(\cdots,\epsilon_jk_j,\epsilon_lk_l,\cdots)
=\frac{\epsilon_jk_j-\epsilon_lk_l+i\,c}{\epsilon_jk_j-\epsilon_lk_l-i\,c}\,
A(\cdots,\epsilon_lk_l,\epsilon_jk_j,\cdots)\,,
\label{eq:kreverse}
\end{equation}
which is from the continuity of $\Psi$ at $x_j=x_{j+1}$:
$c\,\Psi|_{x_{j+1}=x_j}=[\partial\Psi/\partial x_{j+1}-\partial\Psi/\partial x_j]|_{x_{j+1}=x_j}$\,.

The first condition in (\ref{eq:hardwallbc}), $\Psi(x_1=0,x_2,\cdots,x_N)=0$, results in the symmetry of $A$
under the sign reverse of its first argument:
\begin{equation}
A(\epsilon_jk_j,\cdots)=A(-\epsilon_jk_j,\cdots)\,,
\label{eq:signrev}
\end{equation}
and the second condition in (\ref{eq:hardwallbc}), $\Psi(x_1,\cdots,x_{N-1},x_N=L)=0$,
provides the additional conditions on $k$'s:
\begin{equation}
e^{i\,2k_jL}=\frac{A(\cdots,-k_j)}{A(\cdots,k_j)}\,,
\label{eq:Aratio}
\end{equation}
where ``$\cdots$'' abbreviates the same arguments between numerator and denominator.
Apply (\ref{eq:kreverse}) to the numerator of right-hand side in (\ref{eq:Aratio}) iteratively
so that $-k_j$ can be placed in the first argument.
Then its minus sign can be removed due to (\ref{eq:signrev}).
Again apply (\ref{eq:kreverse}) to that numerator repeatedly to get the sign-removed $k_j$ back to its original position,
that is, to the last argument.
Finally, one obtains the Bethe equations from (\ref{eq:Aratio}):
\begin{equation}
e^{i\,2k_jL}=\prod_{l\neq j}\frac{k_j+k_l+i\,c}{k_j+k_l-i\,c}\,\frac{k_j-k_l+i\,c}{k_j-k_l-i\,c}\,.
\label{eq:betheeq}
\end{equation}

Applying the logarithm to the both sides of (\ref{eq:betheeq}) leads to a systems of $N$ real equations
\begin{equation}
Lk_j=\pi n_j+\sum_{l\neq j}\left[\tan^{-1}\Big(\frac{c}{k_j-k_l}\Big)+\tan^{-1}\Big(\frac{c}{k_j+k_l}\Big)\right]
\end{equation}
with integer $n_j$'s satisfying $1\le n_1\le n_2\le\cdots\le n_N$.
A different form of Bethe equations is obtained 
using $\tan^{-1}(X)=\pm\frac{\pi}{2}-\tan^{-1}(\frac{1}{X})$ when $X\gtrless0$:
\begin{equation}
Lk_j=\pi I_j-\sum_{l\neq j}\left[\tan^{-1}\Big(\frac{k_j-k_l}{c}\Big)+\tan^{-1}\Big(\frac{k_j+k_l}{c}\Big)\right],
\end{equation}
where $I_j=n_j+j-1$.
Note that the $n_j$ allow duplicate values, but $I_j$'s do not.
For example, the ground state corresponds to $n_1=n_2=\cdots n_N=1$ or $I_1=1,\,I_2=2,\,\cdots,\,I_N=N$.

The wavefunction (\ref{eq:hardwallsol}) is not normalized yet.
Gaudin conjectured the norm of the Bethe ansatz in the Lieb-Liniger model
under the periodic boundary condition \cite{Gaudin_2014} and Korepin proved it \cite{Korepin_1982}.
For the current hard-wall boundary condition (\ref{eq:hardwallbc}),
we modify Gaudin's conjecture and suggest the normalized form of (\ref{eq:hardwallsol}) as
\begin{equation}
\psi(x_1,\cdots,x_N)=2^{-\frac{N}{2}}\big|\mathbf{H}(B)\big|^{-\frac{1}{2}}\!\!\!\!
\sum_{\epsilon_1,\cdots,\epsilon_N}\!\!\sum_P\epsilon_1\cdots\epsilon_N(-)^P\!
\exp\!\left\{\frac{i}{2}\sum_{j<l}\phi_{P_j,P_l}^{\epsilon_j,\epsilon_l}
-\frac{i}{2}\sum_{j<l}\varphi_{P_j,P_l}^{\epsilon_j,\epsilon_l}
+i\sum_{j=1}^N\epsilon_jk_{P_j}x_j
\right\},
\end{equation}
where $\phi_{P_j,P_l}^{\epsilon_j,\epsilon_l}:=2\tan^{-1}(\frac{c}{\epsilon_jk_{P_j}-\epsilon_lk_{P_l}})$,
$\varphi_{P_j,P_l}^{\epsilon_j,\epsilon_l}:=2\tan^{-1}(\frac{c}{\epsilon_jk_{P_j}+\epsilon_lk_{P_l}})$,
and $(-)^P$ is $+$/$-$1 for even/odd permutation($P$).
Also, the $|\mathbf{H}(B)|$ is the determinant of the Hessian matrix of $B(k_1,\cdots,k_N)$:
\[
B(k_1,\cdots,k_N):=\frac{L}{2}\sum_{j=1}^Nk_j^2-\pi\sum_{j=1}^NI_jk_j\\
+\frac{1}{2}\sum_{j=1}^N\sum_{l\neq j}\Big[\int_0^{k_j-k_l}\!\!\!\!dk\tan^{-1}\!\!\big(\frac{k}{c}\big)
+\!\int_0^{k_j+k_l}\!\!\!\!dk\tan^{-1}\!\!\big(\frac{k}{c}\big)\Big].
\]
Then the norm squared of (\ref{eq:hardwallsol}) can be explicitly written as
\begin{equation}
\mathcal{N}^{\,2}=2^N\prod_{j<l}\Big[1+\frac{c^2}{(k_j-k_l)^2}\Big]\Big[1+\frac{c^2}{(k_j+k_l)^2}\Big]\,
\text{det}\big[\mathbf{H}(B)\big],
\label{eq:hardwallnorm}
\end{equation}
where the entry of the Hessian $\mathbf{H}(B)$ is
\[
\big[\mathbf{H}(B)\big]_{ij}
=\delta_{ij}\bigg[L\,+\sum_{l\neq i}\Big[\frac{c}{(k_i-k_l)^2+c^2}+\frac{c}{(k_i+k_l)^2+c^2}\Big]\bigg]
+\,(1-\delta_{ij})\bigg[-\frac{c}{(k_i-k_j)^2+c^2}+\frac{c}{(k_i+k_j)^2+c^2}\bigg].
\]
We take the $\mathcal{N}$ in (\ref{eq:hardwallnorm}) as the exact norm of (\ref{eq:hardwallsol})
without mathematical rigor being used.
We checked empirically that (\ref{eq:hardwallnorm}) fits well
with the numerical result of
$\int_0^Ldx_N\int_0^{x_N}dx_{N-1}\cdots\int_0^{x_2}dx_1\big|\tilde{\psi}(x_1,\cdots,x_N)\big|^2$. 

\end{widetext}

\nocite{}

\bibliography{FisherLL11}

\end{document}